\documentclass[article,twocolumn,floatfix,superscriptaddress,longbibliography]{revtex4-1}
\usepackage{adjustbox}

\usepackage{amsfonts, amsmath, amssymb}
\usepackage{graphicx}
\usepackage{dcolumn}
\usepackage{mathtools}
\usepackage{bm}
\usepackage[normalem]{ulem}
\usepackage{color}
\usepackage[utf8]{inputenc}
\usepackage[colorlinks,citecolor=blue,linkcolor=blue,urlcolor=blue]{hyperref}
\usepackage{braket}
\usepackage[caption=false]{subfig}
\usepackage{color}
\usepackage{soul}
\usepackage{mathtools}

\usepackage{float}
\newcommand{\be}{\begin{equation}}
\newcommand{\ee}{\end{equation}}
\newcommand{\ba}{\begin{eqnarray}}
\newcommand{\ea}{\end{eqnarray}}

\usepackage{amsfonts, amsmath, amssymb}
\usepackage{graphicx}
\usepackage{dcolumn}
\usepackage{bm}
\usepackage[normalem]{ulem}
\usepackage{color}
\usepackage[utf8]{inputenc}
\usepackage[colorlinks,citecolor=blue,linkcolor=blue,urlcolor=blue]{hyperref}
\usepackage{braket}
\usepackage[caption=false]{subfig}
\usepackage{color}
\usepackage{soul}
\usepackage{mathtools}
\usepackage{float}

\include{physics}

  {\left\lbrace\begin{array}{@{}l@{}}}%
  {\end{array}\right.}

\begin{document}

\author{Nitish Mehta}
\affiliation{Department of Physics, University of Maryland, College Park, MD 20742, USA}
\author{Cristiano Ciuti}
\affiliation{Universit\'{e} Paris Cit\'e, CNRS, Mat\'{e}riaux et Ph\'{e}nom\`{e}nes Quantiques, F-75013 Paris, France}
\author{Roman Kuzmin}
\affiliation{Department of Physics, University of Maryland, College Park, MD 20742, USA}

\author{Vladimir E. Manucharyan}
\affiliation{Department of Physics, University of Maryland, College Park, MD 20742, USA}
\affiliation{Ecole Polytechnique F\'{e}d\'{e}rale de Lausanne (EPFL), CH-1005 Lausanne, Switzerland}

\title{Theory of strong down-conversion in multi-mode cavity and circuit QED}

\begin{abstract}

We revisit the \textit{superstrong} coupling regime of multi-mode cavity quantum electrodynamics (QED), 
defined to occur when the frequency of vacuum Rabi oscillations between the qubit and the nearest cavity mode
exceeds the cavity's free spectral range. A novel prediction is made that the cavity's linear spectrum, measured in the vanishing power limit, can acquire an intricate fine structure associated with the qubit-induced cascades of coherent single-photon down-conversion processes. This many-body effect is hard to capture by a brute-force numerics and 
it is sensitive to the light-matter coupling parameters both in the infra-red and the ultra-violet limits. We focused at the example case of a superconducting fluxonium qubit 
coupled to a long transmission line section. 
The conversion rate in such a circuit QED setup can readily exceed a few MHz, which is plenty to overcome the usual decoherence processes.  Analytical calculations were made possible by an unconventional gauge choice, in which the qubit circuit interacts with radiation via the flux/charge variable in the low-/high-frequency limits, respectively. Our prediction of the fine spectral structure lays the foundation for the “strong down-conversion” regime in quantum optics, in which a single photon excited in a non-linear medium spontaneously down-converts faster than it is absorbed.

\end{abstract}

\date{\today}
\maketitle

\section{Introduction}

Quantum electrodynamics (QED) is a branch of physics describing a remarkable list of fundamental phenomena produced by the quantum nature of electromagnetic fields \cite{Feynman_QED, Milonni_1994,Cohen_Photons}. In cavity QED, the electromagnetic modes are spatially confined and the corresponding vacuum fields can be dramatically enhanced \cite{Haroche2006}. A celebrated manifestation of cavity QED is the \textit{strong} coupling regime, in which an atom (or a qubit) and a cavity mode coherently exchange a single excitation -- undergoing the vacuum Rabi oscillations -- faster than the decoherence rate in either system. This regime has enabled one to control qubits with radiation and to control radiation with qubits, and among other directions it influenced the development of circuit QED and superconducting quantum computing~\cite{ABC_CircuitQED, RMP_circuitQED}.

More recently, two new kinds of strong coupling regimes of cavity QED have been explored. The first one is the \textit{ultrastrong} coupling regime that is achieved when the non-rotating-wave terms of light-matter interaction become relevant, resulting in the non-conservation of the total number of excitations in the system~\cite{Ciuti_2005, Forn_2019,Frisk_Kockum_2019,GarciaVidal2021}. 
In the simplest case of a single mode cavity, ultrastrong coupling physics kicks in when the vacuum Rabi frequency becomes comparable to the atom/cavity transition frequencies, as demonstrated in the semiconductor~\cite{Todorov_2010,Scalari_2012} and the circuit QED~\cite{Niemczyk_2010,Yoshihara_2016} platforms. Another kind of strong coupling regime, termed \textit{superstrong} coupling~\cite{Meiser_2006}, can be obtained in massively multi-mode cavities, when the vacuum Rabi frequency exceeds the free spectral range of the resonator, that is the frequency spacing between the modes. In this case, the qubit exchanges an excitation with the cavity faster than light can traverse the cavity length, and hence the spatial profile of the cavity modes becomes dependent on the qubit state. This regime was approached in circuit QED, using a 
meter-long on-chip superconducting transmission line resonator~\cite{Houck_2015} and in a cold atom setup using a 30 m long optical resonator~\cite{Volz2019}.

The superstrong coupling has an intuitive spectroscopic manifestation in the single-particle approximation. For a cavity with a free spectral range $\Delta$, the qubit resonance at frequency $f_{\rm{eg}} = \omega_{\rm{eg}}/2\pi$ simultaneously hybridizes with about $\Gamma/\Delta$  standing-wave modes of the cavity that are nearby in frequency (Fig.~1b). The quantity $\Gamma$ is, in fact, the rate of spontaneous emission of the qubit in the limit of an infinitely long cavity, corresponding to $\Delta \rightarrow 0$, and the superstrong coupling condition formally reads $\Delta \ll \Gamma \ll f_{\rm{eg}}$. Because there are many cavity modes and only one qubit, one can think that each cavity mode from the $\Gamma$-vicinity of the qubit resonance becomes weakly dressed by the qubit. The dressed modes acquire a small frequency shift, of the order $\Delta^2/\Gamma$, which rapidly vanishes outside the hybridization window. This spectral property was indeed verified in recent cQED experiments, where the ratio $\Gamma/\Delta$ was increased further compared to the Ref.~\cite{Houck_2015} using compact high-impedance/slow light transmission lines~\cite{ Puertas_Mart_nez_2019,Kuzmin_2019}.

\begin{figure}
    \centering
    \includegraphics[width =0.5\textwidth]{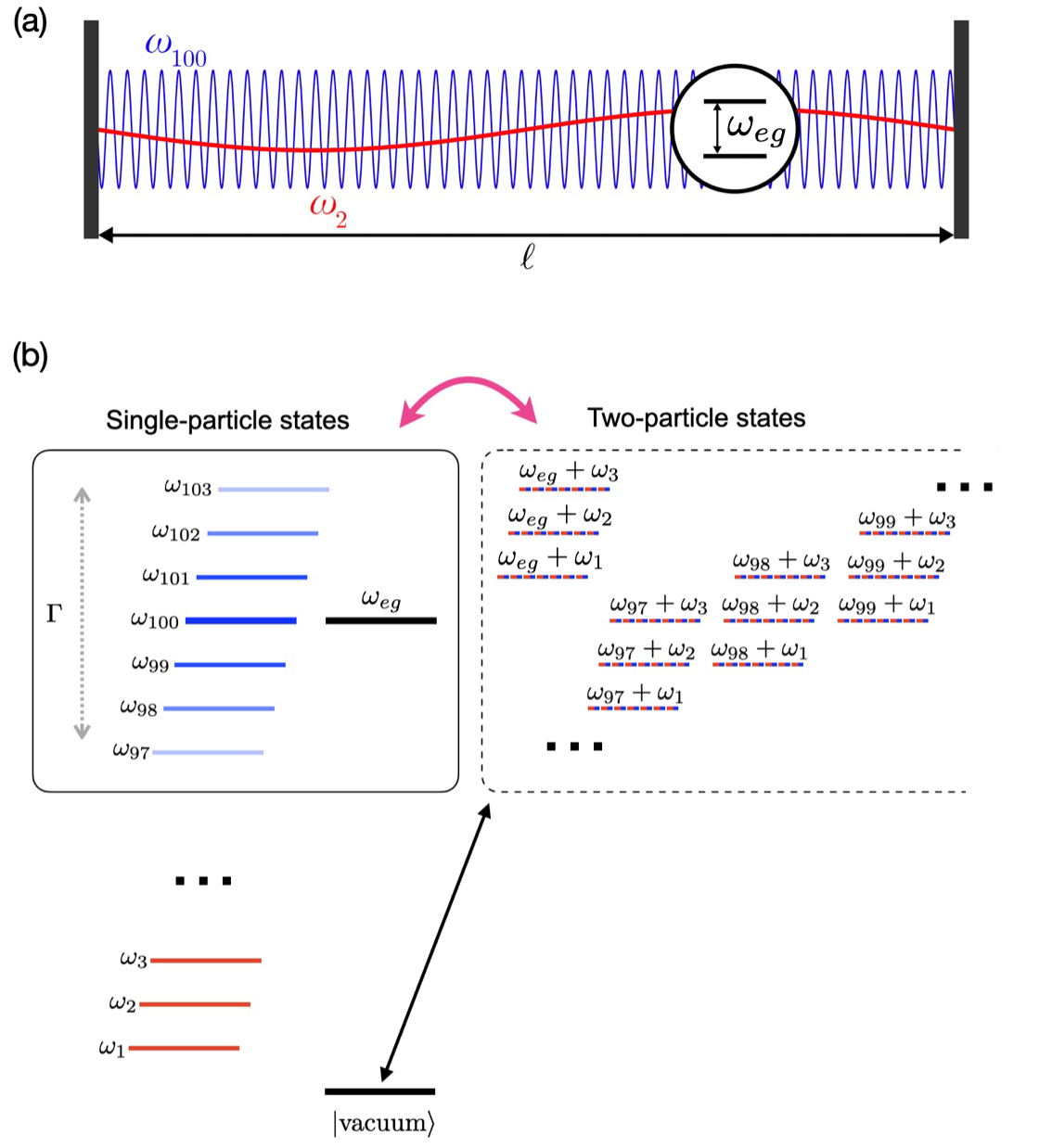}
    \caption{(a) Schematic of modes in an ideal Fabry-Perot cavity resonator coupled to a single atom (qubit) at frequency $\omega_{eg}$ The bare mode frequencies are equally spaced by frequency $\Delta$ and are labeled $\omega_{1}$, $\omega_{2}$, ..., $\omega_{100}$, ...  (b, left side) Single-particle states of the uncoupled system, separated into one photon in the low-frequency modes (red), one photon in the high-frequency modes (blue), as well as one qubit excitation, labeled $\omega_{\textrm{eg}}$ (black). In the superstrong coupling regime, a number of high-frequency modes in the $\Gamma$-vicinity of the qubit transition hybridize with the qubit. (b, right) Examples of two-particle excitations in the same energy window, involving one photon in the cavity's lowest-frequency modes. For concreteness, we set $\omega_{\textrm{eg}} = \omega_{100}$.
    \label{intro-schematic}}
\end{figure}

Here we reveal a surprisingly important role of the cavity's lowest energy modes in the superstrong coupling dynamics (Fig~\ref{intro-schematic}a,b). Even though these far-detuned modes are negligibly dressed by the qubit, one can use them to construct a large number of multi-particle excitations with energies near the qubit resonance. For example, in addition to a number around $\Gamma/\Delta$ of single-particle states hybridizing with the qubit, there is a much larger number of two-particle states, scaling as $(\Gamma/\Delta)^2$, in the same energy window. These states consist of one ``high-frequency" photon at a frequency near $f_{\rm{eg}}$ and one ``low-frequency" photon at a frequency of about $\Gamma$ or less. There is an even larger number of three-particle states, involving two low-frequency photons in the cavity, and so forth. The presence of low-frequency modes is crucial for such a massive multi-particle degeneracy: if we ignore the cavity's spectrum below half the qubit frequency, there would be room for only one two-particle excitation and no three-particle ones. How strong can the coupling between single-particle and multi-particle manifolds be in the superstrong coupling regime, and how would this many-body interaction effect manifest in the simplest \textit{linear} (vanishing probe power) spectroscopy experiment? Our work  comprehensively answers this question.

To understand the cavity's multi-particle dynamics, one has to first face the notorious problem of the gauge choice in quantum electrodynamics (see e.g. the references \cite{Bassani_1977,Todorov_2012,Manucharyan_2017, De_Bernardis_gauge,Di_Stefano_2019,stokes2019gauge,Di_Vincenzo_2019,Nazir2020,Nazir2020b,dmytruk2021gauge,Savasta2021,Nazir2021,Savasta2022}). Indeed, a coherent hybridization between a single-particle and, say, a two-particle excitation in Fig~\ref{intro-schematic}b clearly violates the conservation of the total number of excitations, which is the key attribute of the ultrastrong coupling regime. However, it is not obvious which Hamiltonian parameters control the emergence of ultrastrong coupling physics in a massively multi-mode system. To make matters worse, the light-matter coupling Hamiltonians are generally gauge-dependent~\cite{Manucharyan_2017}, which in the past has led to paradoxical predictions for multi-mode problems, such as a divergent Lamb shift or spontaneous emission rate~\cite{Steele_2017, Tureci_2017, Tureci2022}. 
Furthermore, simulations of multi-mode QED systems are impossible without truncating the system's Hilbert space, and the accuracy of such a truncation is also gauge-dependent, as it was first pointed out for the problem of a two-photon absorption in hydrogen \cite{Bassani_1977}. To our knowledge, no general recipe exists for choosing the most computationally efficient gauge for a multi-mode cavity QED problem \cite{Di_Vincenzo_2019}.

In the circuit QED platform, particularly suited for exploring quantum effects beyond the strong coupling, the majority of multi-mode experiments involved a transmon type qubit, which is in fact a weakly anharmonic oscillator circuit rather than a two-level system~\cite{koch2007charge}. In this case, the black box circuit quantization (BBQ) theory provides an efficient solution to all challenges related to the gauge choice \cite{manucharyan2007microwave, Nigg_2012}. Within the BBQ theory, the qubit's non-linearity is treated as a perturbation to the normal modes of the coupled system, the parameters of which are given entirely in terms of the circuit impedance or admittance functions. However, the conversion processes are suppressed, at least at the single-photon level, because the interaction felt by the low-frequency modes is controlled entirely by the degree of their hybridization with the qubit, a vanishing quantity in the superstrong coupling regime. Here we consider the case of a highly anharmonic fluxonium qubit~\cite{manucharyan2009fluxonium, manucharyan2012evidence},
to which the BBQ theory cannot be directly applied.

Our new result can be summarized as follows. The single-particle excitations of the \textit{superstrongly} coupled qubit-cavity system can be formally viewed as polaritons, although we prefer to call them ``dressed photons", because the qubit degree of freedom is also of the electromagnetic nature. These dressed photons experience a non-linearity, the dominant effect of which is to convert a single dressed photon into a lower energy dressed photon and a number of low-frequency photons (the dressing of which is negligible). We focused at the most efficient conversion process, involving just one low-frequency photon production. Such a parity-flipping process is allowed by symmetry when operating the qubit circuit away from a half-integer flux bias. In this case, the single-photon conversion rate can readily reach a few MHz, which is well above the decoherence floor in superconducting circuits, and which makes the conversion process \textit{reversible} and \textit{coherent}. Consequently, the cavity's linear spectrum of standing-wave modes would acquire a fine structure of multi-particle resonances, consisting of hybridized single-particle and two-particle excitations. Higher-order conversion processes can lead to even finer spectral features, involving three-particle states and so forth.

Our prediction 
has important connections with two other topics. The first connection is the quantum impurity dynamics of the Ohmic bath spin-boson model~\cite{Girvin_2019, forn2017ultrastrong}. The hybridization of single-particle and multi-particle excitations would lead, in the infinite cavity size limit, to a non-zero probability of the inelastic scattering of a single photon by the qubit~\cite{Goldstein_2013, Florens_2018, houzet2020critical, burshtein2021photon}. 
The second connection is to the phenomenon of spontaneous parametric down-conversion (SPDC)~\cite{klyshko1969scattering, harris1967observation}. To our knowledge, there hardly exist a non-linear optical material where the SPDC rate would be larger or at least comparable to the absorption rate~\cite{yanagimoto2022temporal}. 
Consequently, the probability of observing a reversible SPDC process is practically zero. Our synthetic system implements the opposite scenario: a single photon can coherently down-convert into a photon pair, or even to a superposition of multiple photon pairs, and then up-convert back into the original single photon, and so on, at a sub-microsecond time scale. By analogy with the vacuum Rabi oscillations in cavity QED, we 
refer to this previously unavailable regime of non-linear optics as a regime of \textit{strong down-conversion}. 

The article is organized as follows. In Sec. \ref{section:system}, we present the proposed circuit implementation of multi-mode cavity QED
and describe its Hamiltonians in three gauges, namely the charge gauge, the flux gauge and a newly introduced mixed gauge. The mixed gauge is our key innovation as it provides a shortcut to calculating the two-particle amplitudes. In Sec. \ref{section:effective}, we describe an effective photon-photon interaction Hamiltonian responsible for coherent conversion processes. An example of the resulting fine-structured cavity spectrum is given in Section \ref{section:fine}. In Section \ref{section:conclusions} we draw our conclusions and perspectives. Technical details are organized in several Appendices. In particular, in Appendix \ref{section:benchmark} we present a successful numerical benchmark of the proposed effective Hamiltonian with a direct numerical diagonalization, which was made possible by considering a much shorter-length transmission line and a weaker qubit coupling.

\section{circuit model and gauge choice}
\label{section:system}
\begin{figure}[t!]
    \centering
    \includegraphics[width =0.5\textwidth]{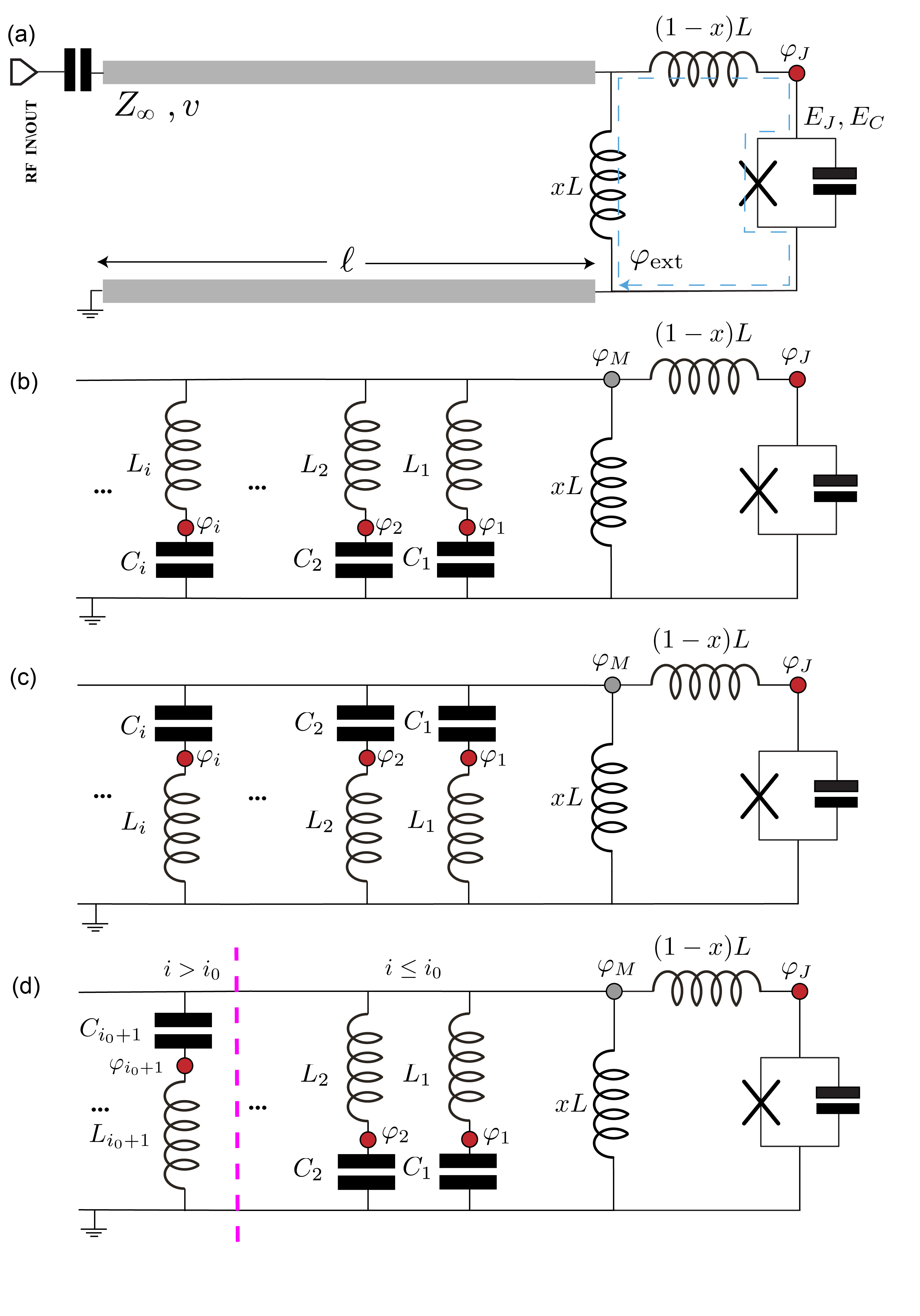}
    \caption{(a) Schematic of a telegraph transmission line section terminated by a superconducting fluxonium qubit. The transmission line is characterized by its wave impedance $Z_{\infty}$, speed of light $v$, and length $\ell$. The fluxonium circuit is defined by the Josephson energy $E_J$, the charging energy $E_C$ of the small junction and the loop inductance $L$. A fraction $x$ of the fluxonium loop inductance is shared with the transmission line.
    The qubit transition frequency can be tuned by changing the external flux bias $(\hbar/2e) \varphi_{\mathrm{ext}}$ threading the superconducting loop.  (b), (c) and (d):  Exact lumped-element circuit model of the device obtained using Foster's decomposition. 
    The values $C_i$ and $L_i$ are derived in the Appendix \ref{foster_appendix}. The filled red dots indicate the choice of generalized coordinate variables $\varphi_i$ corresponding to (b) flux gauge, (c) charge gauge, and (d) mixed gauge. Note that the variable $\varphi_M$ is dependent on $\varphi_i$ and $\varphi_J$ via current conservation.}
    \label{circuit_model}
\end{figure}

In our system, the multi-mode cavity is implemented as a superconducting transmission line of length $\ell$, characterized by the wave impedance $Z_{\infty}$ and the speed of light $v$, as shown in Fig.~\ref{circuit_model} (a). This transmission line is terminated on one end by a superconducting loop of inductance $L$, interrupted by a single weak Josephson junction, characterized by the Josephson energy $E_J$ and the charging energy $E_C$. With the proper choice of parameters, such a loop implements a circuit atom known as fluxonium. Fluxonium interacts with the transmission line modes by sharing a fraction $x$ ($0 < x \leq 1$) of its loop inductance with the line. Fluxonium's spectral properties can be tuned by varying an external magnetic flux $(\hbar/2e)\varphi_{\rm ext}$ pierced through the superconducting loop. The hybridization parameter $\Gamma$ grows with $x$ and it also depends on $Z_{\infty}$ and the fluxonium circuit parameters. All numerical calculations in the main text are done using example device parameters listed in Table~\ref{tableApp}.

The atom-cavity coupling Hamiltonian for the system shown in Fig. \ref{circuit_model} (a) takes the usual form:

\begin{equation}
    \hat{H} = \hat{H}_{\textrm{atom}} + \hat{H}_{\textrm{modes}} + \hat{H}_{\textrm{int}},
    \label{H_tot}
\end{equation}
where the first two terms describe respectively an atom and a set of non-interacting bosonic modes hosted by the transmission line, while the third term describes their interaction. In the limit $x\rightarrow 0^+$, the fluxonium atom decouples from the transmission line modes. The bare atom Hamiltonian is then given in terms of the phase-difference operator $\hat{\varphi}_J$ across the junction and the conjugate Cooper pair number operator $\hat{n}_J$, which obey the commutation relation $[\hat{\varphi}_J, \hat{n}_J] = \mathrm{i}$. Such Hamiltonian reads:
\begin{equation}
        {\hat {H}}_{\textrm{atom}} = 4E_C {\hat{n}}^2_J - E_J \cos ({\hat{\varphi}}_J - \varphi_{\mathrm{ext}} )
   + E_L {\hat{\varphi}}^2_J/2,
 \label{fluxonium}
\end{equation}
where $E_J$ is the Josephson junction's energy, $E_C = e^2/2C_J$ is the charging energy, $E_L = (\hbar/2e)^2/L$ is the inductive energy of the loop inductance, and  $\varphi_{\textrm{ext}}$ is the external phase-bias, created by piercing the fluxonium loop with an externally applied magnetic field. 

In order to obtain the Hamiltonian (\ref{H_tot}) for $x >0$, as the first step, we replace the continuous electromagnetic structure of the transmission line by its lumped circuit element representation. The lumped element model involves $N$ series-$LC$ circuits connected in parallel. The values of the individual capacitors $C_i$ and inductors $L_i$ ($i=1,2,...,N$) can be obtained using Foster's theorem (see Appendix ~\ref{foster_appendix}), which states that for $N \rightarrow \infty$ the lumped element representation becomes mathematically exact. To simplify the discussion, we consider a non-dispersive, open-ended transmission line, characterized by only two parameters: the wave impedance $Z_{\infty}$, and the free spectral range $\Delta = v/2\ell$, where $v$ is the speed of light and $\ell$ is the transmission line length. 

In the second step, we must choose an electromagnetic gauge, that is a set of independent generalized coordinates of the circuit. Depending on the gauge, the circuit atom can be coupled to the modes via the flux variable, charge variable, or both. We will show that both the pure flux and pure charge gauge present challenges for truncating either the cavity modes or the atomic levels, and introduce a special mixed gauge which resolves the truncation problem.

\subsection{Flux gauge}

\noindent One common option is to choose the generalized coordinate variables $\varphi_i$ such that they represent the phase-difference across the capacitors $C_i$ (see Fig. \ref{circuit_model} (b), red nodes). Note that the phase $\varphi_M$ is linked to $\varphi_i$ and $\varphi_J$ via the current conservation at the node $M$ (see Fig.~\ref{circuit_model}b, gray node). Following the standard quantization procedure detailed in Appendix \ref{hybrid_appendix}, we obtain the following Hamiltonian operators:
\begin{equation}
    \hat{H}_{\textrm{modes}} = \sum_{i = 1}^{N} \hbar\omega_i^{(f)} \hat{b}^{\dagger}_{i} \hat{b}_{i},
\end{equation}

\begin{equation}
    \hat{H}_{\textrm{int}} =  -\hat{\varphi}_J\sum_{i=1}^{N} h g_i^{(f)} (\hat{b}_i  + \hat{b}^{\dagger}_i).
\end{equation}
Here $\hat{b}_i$ ($\hat{b}_i^{\dagger}$) is the photon annihilation (creation) operator for the photon in the mode $i$ of the transmission line, and the mode frequencies $\omega_i^{(f)}$, calculated through the procedure described in Appendix \ref{hybrid_appendix}, are numerically close to the values $(L_i C_i)^{-1/2}$. The expression for the coupling constants $g_i^{(f)}$ is given by Eq.~(\ref{couplings_flux}). Since the coupling is achieved via the flux variable $\varphi_J$, we call the present variable choice the ``flux" gauge. Importantly, in the flux gauge, the atomic part $\hat{H}_{\textrm{atom}}$ is modified from the bare form of Eq.~(\ref{fluxonium}) by the replacement (see Appendix \ref{hybrid_appendix}):
\begin{equation}
E_L \rightarrow \tilde{E}_L,
\end{equation}
where the renormalized inductive energy $\tilde{E}_L$ is given by Eq.~(\ref{renormalized_L}). Note that for $x=0$ we have $\tilde{E}_L = E_L$, but for $x>0$ the two quantities can differ significantly. 

\begin{figure*}[t!]
    \centering
    \includegraphics[width =1.0\textwidth]{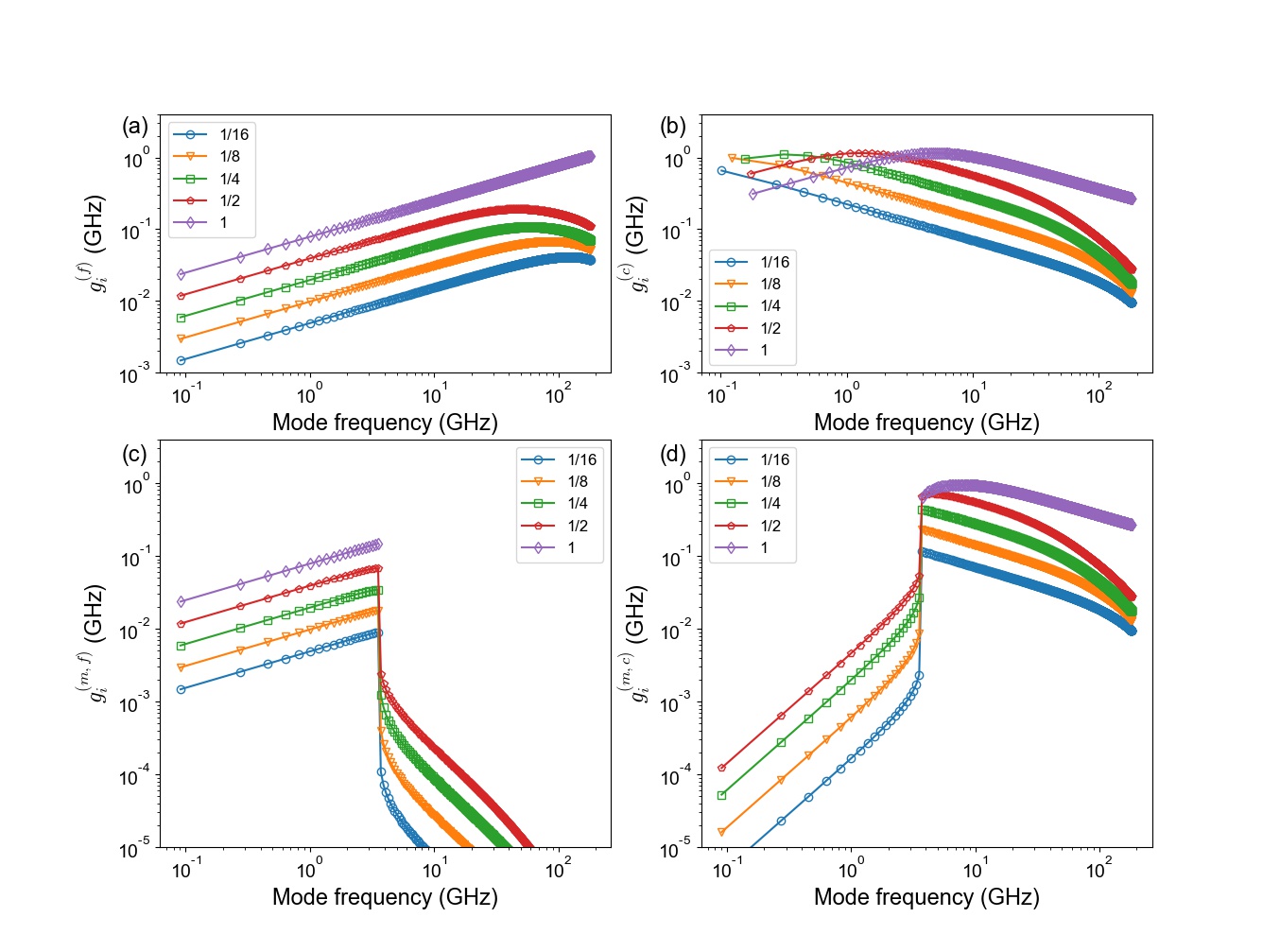}
    \caption{Calculated couplings between atom and transmission line modes in the flux gauge (a), charge gauge (b), and the mixed gauge, for $i_0 = 20$ (c, d) for several values of the inductive coupling ratio $x$. Other parameters are given in Table~\ref{tableApp}. 
    Note, the coupling constants calculated in the flux gauge scale as $g^{(f)}_i \sim i^{1/2}$ in the large $i$ limit and diverge as $i \to + \infty$. On the other hand, the coupling constants calculated in the charge gauge scale as $g^{(c)}_i \sim i^{-1/2}$ in the large $i$ limit and vanish as $i \to +\infty$. The mixed gauge parameters $g^{(m,f)}_i$ and $g^{(m,c)}_i$ interpolate the properties of the two pure gauges.
    }
    \label{couplings}
\end{figure*}

Unfortunately, in the flux gauge the separation of Hamiltonian (\ref{H_tot}) into the three terms has a number of disturbing properties. To start, the value of the coupling constant $g_i^{(f)}$ grows with $i$ approximately as $i^{1/2}$ (see Fig.~\ref{couplings} (a)). Therefore, truncating the model to a finite number of low-energy modes cannot be readily justified. 
In addition, the renormalized inductive energy $\tilde{E}_L$ also depends on the total number of modes used in the Foster's decomposition of the transmission line ($N$) and can become much larger than $E_L$ for $1-x \ll 1$ (see Fig. \ref{flux_renorm}). In fact, for $x \to 1^-$, the parameter $\tilde{E}_L$ diverges as $\tilde{E}_L \rightarrow E_L/(1-x)$ for $N \rightarrow + \infty$. Such a behavior of $\tilde{E}_L$ is problematic for interpreting the dynamics of the renormalized atom's Hamiltonian, because the two lowest energy eigenstates of $\hat{H}_{\textrm{atom}}$ loose their spin-like nature for $E_L > E_J$ (the Josephson potential looses its degenerate local minima at $\varphi_{\textrm{ext}} = \pi$). For the reasons above, it is challenging to produce even qualitative predictions for the outcome of a simple spectroscopy experiment in the single-photon excitation regime.

\subsection{Charge gauge}

We now show that the problem of divergent coupling constants in the ultraviolet limit along with the accompanying $N$-dependence of the model parameters can be eliminated simply by switching to the charge gauge. Specifically, we consider a dual choice of circuit variables shown in Fig.~\ref{circuit_model} (c), where $\varphi_i$ is the phase-difference across the inductance $L_i$. The junction variable $\varphi_J$ remains the same as in the flux gauge, and the dependent variable $\varphi_M$ is still given by the current conservation at the node $M$. With the new choice of variables, the atom couples to the bosonic annihilation (creation) operators $\hat{b}_i$ ($\hat{b}^{\dagger}_i$) of the transmission line modes via the Cooper pair number operator $\hat{n}_J$. The transmission line and interaction terms of the Hamiltonian read:
\begin{equation}
    \hat{H}_{\textrm{modes}} = \sum_{i = 1}^{N} \hbar \omega_i^{(c)} \hat{b}^{\dagger}_{i} \hat{b}_{i},
\end{equation}

\begin{equation}
    \hat{H}_{\textrm{int}} =  \mathrm{i} \hat{n}_J\sum_{i=1}^{N} h g_i^{(c)} (\hat{b}_i  - \hat{b}_i^{\dagger}) \\.
\end{equation}

Importantly, the parameters in $\hat{H}_{\textrm{atom}}$ exactly match their bare ($x=0$) values in the charge gauge. The mode frequencies $\omega_i^{(c)}$, calculated  in Appendix \ref{hybrid_appendix}, are also numerically close to the values $(L_i C_i)^{-1/2}$. The coupling constants $g_i^{(c)}$, as shown in Appendix \ref{hybrid_appendix}  and Eq.~(\ref{couplings_charge}), are independent of $N$ for $i < N$. Most importantly, while the dependence of the coupling constants $g_i^{(c)}$ vs. $i$ on the circuit parameters $x$ and $Z_{\infty}$ can be quite complex (Fig.~\ref{couplings} (b)), the coupling necessarily drops faster than $i^{-1/2}$ for $i \gg 1$. This property of the charge gauge guarantees the convergence of perturbative calculations of quantities such as the Lamb shift or Purcell effect and justifies the truncation of the cavity's high-energy modes.

Nevertheless, the charge gauge introduces a different profound problem. Note that the expectation value of the Cooper pair number operator $n_J$ is zero independently of the flux bias, and hence the qubit transition appears to obey the inversion symmetry at any flux bias. Consequently, the symmetry considerations forbid the coupling to two-particle states in the charge gauge. The situation is different in the flux gauge, because there the qubit is coupled via the $\varphi_J$ variable, the expectation value of which is indeed non-zero, and state-dependent, when the circuit is biased away from the flux sweet-spots. The resolution for this apparent paradox lies in the fact that one simply cannot truncate the atomic spectrum to the two lowest qubit levels in the charge gauge. A similar failure of the two-level approximation occurs for a two-photon absorption in hydrogen in the Coulomb gauge~\cite{Bassani1977}. Thus, while the charge gauge is convenient for truncating the cavity modes, it requires taking into account potentially many atomic levels, which is also a major inconvenience for both the physical transparency and numerical calculations.

\subsection{Mixed gauge}
Our solution to the pure flux/charge gauge issues is to introduce a mixed gauge, which would combine the best properties of the two pure gauges. Specifically, we consider the choice of circuit variables shown in Fig. \ref{circuit_model} (d), where $\varphi_i$ is the phase-difference across the capacitance $C_i$ for $i \leq i_{0}$, while for $i > i_{0}$ it is the phase difference across the inductance $L_i$.  The junction variable $\varphi_J$ remains the same as in the flux and charge gauge, and the dependent variable $\varphi_M$ is still given by the current conservation at the node $M$. In the following, we will call this new choice of variables the “mixed” gauge since the resulting Hamiltonian features the atom variables coupled to the bosonic annihilation (creation) operators $\hat{b}_i (\hat{b}^{\dagger}_i)$ of the transmission line modes via both the Cooper pair number operator $\hat{n}_J$ and the phase-difference operator $\hat{\varphi}_J$.  The transmission line and interaction terms of the Hamiltonian read:
\begin{equation}
    \hat{H}_{\textrm{modes}} = \sum_{i = 1}^{N} \hbar \omega_i^{(m)} \hat{b}^{\dagger}_{i} \hat{b}_{i},
\end{equation}

\begin{equation}
    \hat{H}_{\textrm{int}} = -\hat{\varphi}_J \sum_{i=1}^{N} hg^{(m,f)}_i (\hat{b}_i + \hat{b}^{\dagger}_i) - \mathrm{i} \hat{n}_J\sum_{i=1}^{N} h g_i^{(m,c)} (\hat{b}_i  - \hat{b}_i^{\dagger}).
\end{equation}

Note that the notation $(m)$ stands for mixed gauge, which also implicitly depends on the parameter $i_0$. Since this mixed gauge interpolates between pure charge and flux gauges, it will inherit properties from both gauges as the parameter $i_{0}$ is changed (see Fig. \ref{couplings} (c, d), where $i_0 = 20$). In the special case of $x=1$, one can see that $g^{(m,c)}_i = 0$ for $i \leq i_0$ and $g^{(m,f)}_i = 0$ for $i > i_0$, i.e. the system is coupled via the charge variable to the high-frequency modes and via the flux variable to the low-frequency modes. For $x < 1$, there is some residual coupling via charge/flux variable at low/high frequencies, but it is in fact quite negligible. Likewise, there is a residual flux-gauge renormalization of the $E_L$ parameter left, but it is cut off at $i_0$ and, for example, for $i_0 =20$, the effect is quite small and can be taken into account phenomenologically (see Fig.~\ref{flux_renorm}). 

The proposed mixed gauge choice is favorable because (i) it justifies the truncation of the high-frequency cavity modes just like the charge gauge and (ii) it allows for the parity non-conserving conversion processes even if the qubit circuit Hamiltonian is truncated to its two lowest energy levels, as the low-frequency modes are now ``aware" of the parity symmetry breaking by the external flux bias.

\begin{figure*}[t!]
    \centering
    \includegraphics[width =0.8\linewidth]{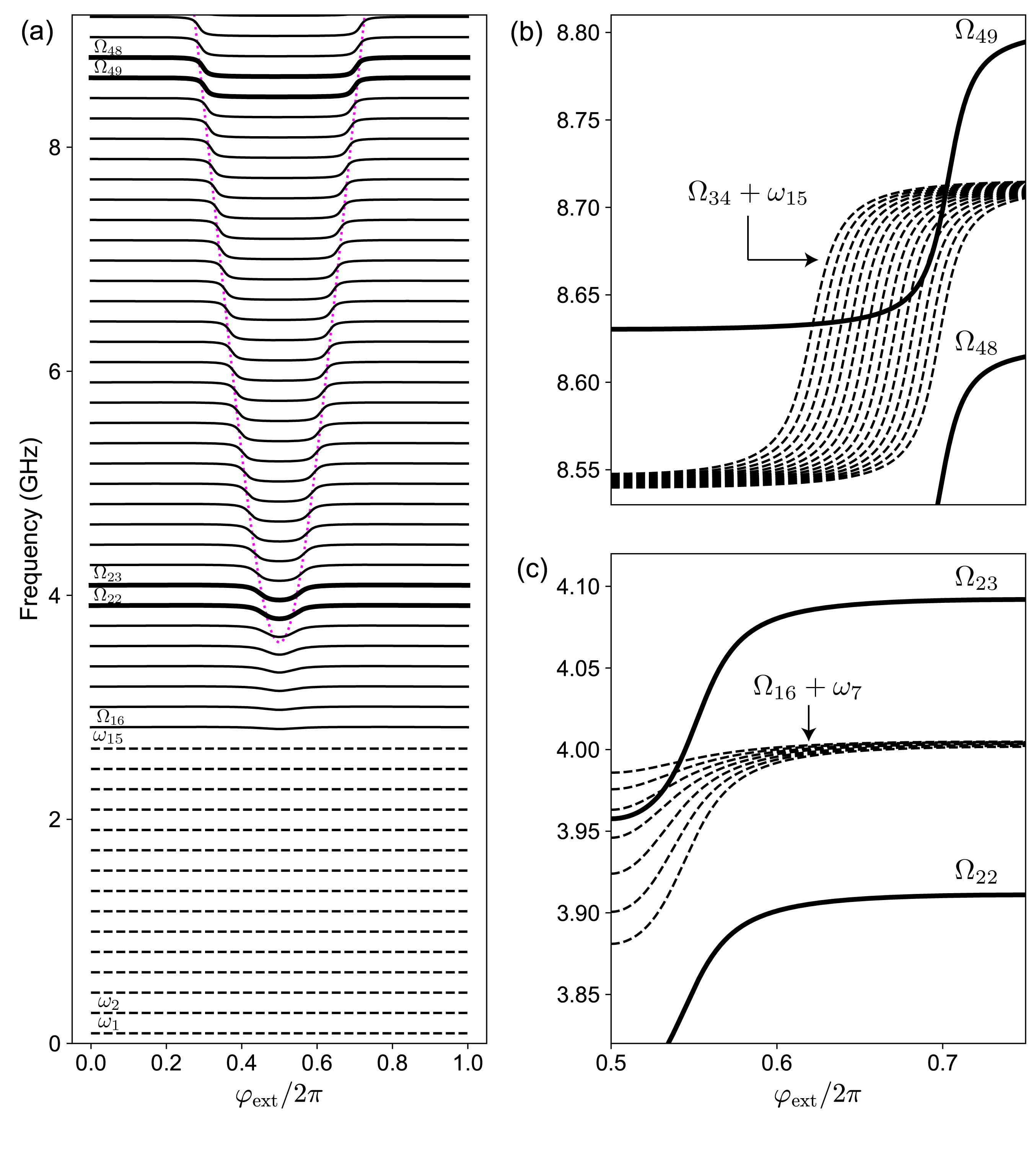}
    \caption{(a) Single-particle spectrum as a function of the flux tuning parameter $\varphi_{\textrm{ext}}$ for a system consisting of $N=250$ equally spaced modes (only the lowest 50 levels are shown). The undressed lowest-frequency modes are shown by dashed lines, the separator index is $i_0 = 15$. The bare qubit frequency $f_{eg}(\varphi_{\textrm{ext}})$ is indicated with a dotted magenta line. (b), (c): Close-ups near polariton  branches $k = 23$ and $k = 49$, showing examples of bare two-photon energies in the same energy window. Multi-particle interaction is not yet taken into account in this plot.}
    \label{long_chain}
\end{figure*}

\section{Effective multi-particle Hamiltonian}
\label{section:effective}

In this section we derive the effective photon-photon interaction induced by the qubit using the mixed gauge. We truncate the fluxonium part of the Hamiltonian to its lowest two energy states, connected by a flux-tunable qubit transition at frequency $f_{\rm eg}(\varphi_{\textrm{ext}})$. In what follows, our treatment is valid as long as $\Gamma \ll f_{\rm eg}$, and the default choice for $i_0$ is $(f_{\rm{eg}}/\Delta)/2$. In a specific numerical calculation it might make sense to use a slightly lower cutoff for a better accuracy.

The procedure for obtaining the effective Hamiltonian that will be discussed here is detailed in Appendix \ref{section:analytical}. 
The first step is to introduce for the high-frequency modes $(k > i_{0})$ the ``polaritonic"  (dressed photon) excitations. These are linear superpositions of the qubit excited state with the cavity's single-photon states, as described by:
\begin{equation}
    \hat{p}^{\dagger}_k \vert G \rangle = W_{k,0} \vert \mathrm{e} \rangle \vert 0 \rangle + \sum_{k'> i_{0}}  W_{k, k'} \vert \mathrm{g} \rangle \hat{b}^{\dagger}_{k'} \vert 0 \rangle \, ,
\end{equation}
where $\vert G \rangle = \vert \mathrm{g} \rangle \otimes \vert 0 \rangle $ is the ground state of the system (fluxonium in the ground state and zero excitations in the transmission line modes). The hybridization coefficients and corresponding energies can be determined by diagonalizing the full Hamiltonian in the subspace spanned by the states $\vert \mathrm{e} \rangle \vert 0 \rangle$ and $\vert \mathrm{g} \rangle \hat{b}^{\dagger}_{k'} \vert 0 \rangle$ for $k' > i_0$. The resulting single-polariton frequencies $\Omega_{k}/2 \pi$ are shown in solid black lines in Fig. \ref{long_chain} (a) for an illustrative set of parameters. The vanishingly small flux dispersion of polaritons near the low frequency cutoff justifies the approximation of ignoring the coupling of the qubit to the low-frequency photonic modes. The single-particle spectrum in Fig. \ref{long_chain} (a) matches the original theoretical predictions \cite{Meiser_2006} and experimental observations \cite{Puertas_Mart_nez_2019}\cite{Kuzmin_2019} made about the superstrong coupling regime.

 However, so far we have completely ignored the effect of the low-frequency modes and restricted the total number of excitations in the system to one. The next step is to enlarge the Hilbert space by including two-particle states consisting of one polariton (dressed photon) excitation and one (undressed) photon occupying one of the cavity's low-frequency modes. The effective photon-photon interaction Hamiltonian, which does not conserve the number of particles is given by:
\begin{widetext}
\begin{equation}
    \hat{H}_{\mathrm{eff}} =  \sum_{i = 1}^{i_{0}} \hbar \omega^{(m)}_i \hat{b}^{\dagger}_i \hat{b}_i +  \sum_{k> i_{0}} \hbar \Omega_k \hat{p}^{\dagger}_k \hat{p}_k + hg \sum_{i=1}^{i_{0}} \sqrt{i}  \sum_{k,k' > i_{0}}  A_{k, k'} (\hat{p}_k \hat{p}^{\dagger}_{k'} \hat{b}^{\dagger}_i + \hat{p}^{\dagger}_k \hat{p}_{k'} \hat{b}_i) \, , \label{effective_Ham}
\end{equation}
\end{widetext}
where
\begin{equation}
    A_{k, k'} =  W^{*}_{k,0} W_{k', 0}  \, , \label{Akk}
\end{equation}
\begin{equation}
    g = g^{(m,f)}_1 (\langle \mathrm{e} \vert \hat{\varphi}_J \vert \mathrm{e} \rangle -  \langle \mathrm{g} \vert \hat{\varphi}_J \vert \mathrm{g} \rangle) \, \label{g} .
\end{equation}

The photon-photon interaction term in Eq.~\ref{effective_Ham} is equivalent to a three-wave mixing, but in the absence of any strong classical drive. The interaction connects $n$-particle states to $(n\pm 1)$-particle states with the following strict selection rule: (i) both states consist of one polaritonic excitation and the rest are the low-frequency photons (formally defined by photons occupying the undressed modes with $i \leq i_0$) and (ii) both states must have the same configuration of low-frequency photons except for one. A softer selection rule requires that the polaritonic excitations in the two coupled states have a frequency difference not much higher than $\Gamma$, i.e. $|\Omega_k - \Omega_{k'}| \lesssim \Gamma$, as can be seen from plotting the matrix $A_{k,k'}$ (see Fig.~\ref{qubit_comp}). 

There are two more profound limitations on the interaction term. First, the interaction is strictly zero at $\varphi_{\mathrm{ext}} = \pi$, at which point the fluxonium potential energy has an inversion symmetry. Away from this flux sweet-spot, the interaction is controlled by the symmetry breaking factor $\langle \mathrm{e} \vert \hat{\varphi}_J \vert \mathrm{e} \rangle -  \langle \mathrm{g} \vert \hat{\varphi}_J \vert \mathrm{g} \rangle$, the maximal value of which can be estimated as $\pi$, the distance between the two fluxonium potential wells. Second, our approximation yields no direct coupling to three-particle states. In order to obtain this effect one must take into account the qubit dressing of the lowest frequency modes, which should become relevant at $\Gamma \approx f_{eg}$, where our model fails and the system enters the conventional ultrastrong coupling regime.

\begin{figure*}
    \centering
    \includegraphics[width =0.85\textwidth]{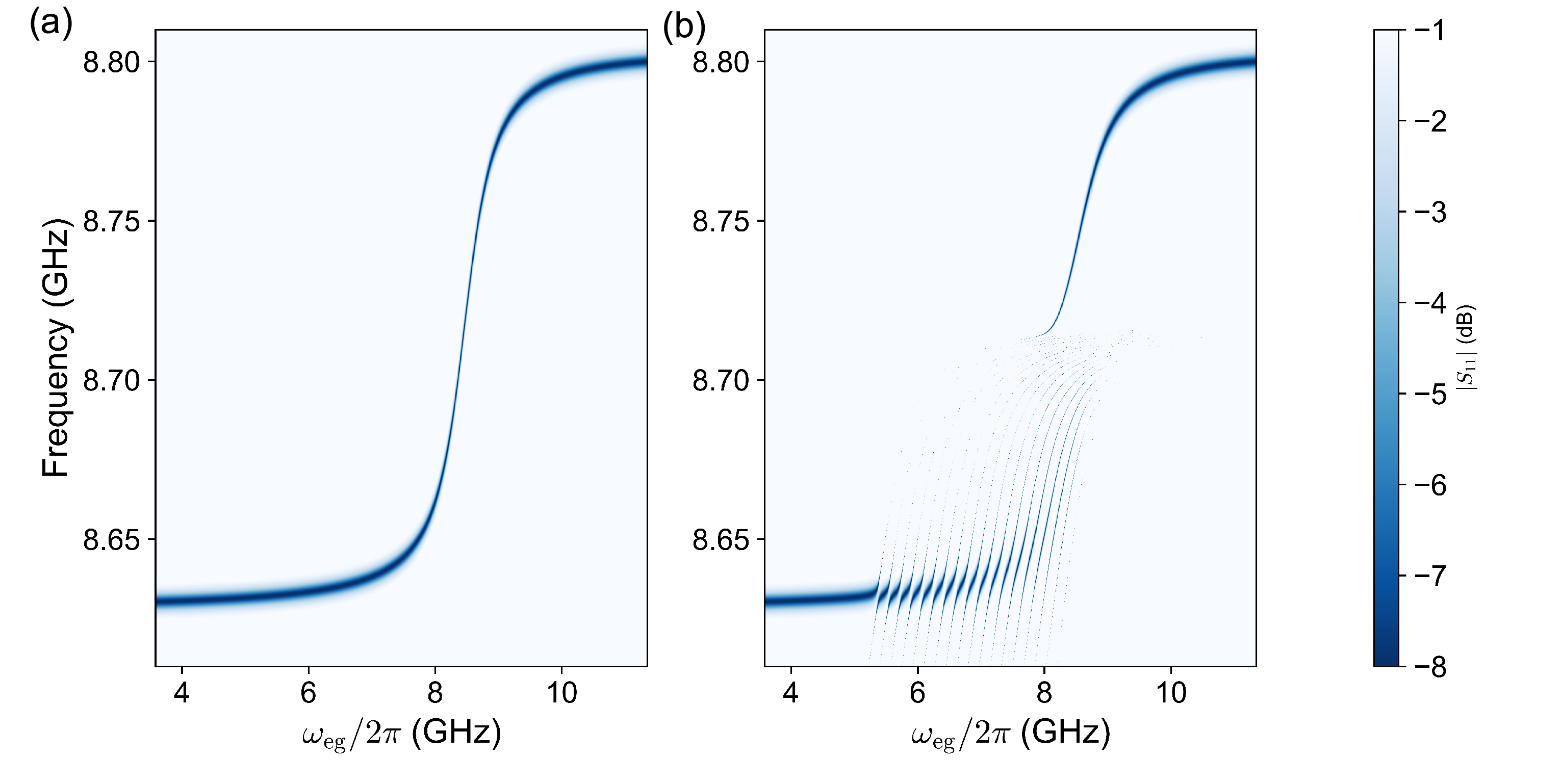}
    \caption{(a) Linear reflection magnitude $|S_{11}|$ in our multi-mode QED setup versus the bare qubit frequency $\omega_{\mathrm{eg}} /2\pi$ (tuned by flux) and probe frequency, near mode $k=49$, calculated in the single-particle approximation. See text for details. The result agrees with the standard picture of superstrong coupling~\cite{Meiser_2006}.
    (b) Same as (a) but calculated in the two-particle approximation. 
    \label{S11}}
\end{figure*}

Let us discuss the maximal conversion rate, i.e. the maximal value of the interaction matrix element in Eq.~\ref{effective_Ham}. The elements of the matrix $A_{k, k'}$ are proportional to the product of the qubit components in the involved polariton modes via the factor $W^{*}_{k,0} W_{k', 0}$. Its maximal value is given approximately by the normalization factor $\Delta/\Gamma$. The vacuum Rabi coupling for the lowest mode is given by $g_1^{(m,f)} \approx g_1^{(f)} = (x \Delta/4)\sqrt{2 R_Q/\pi Z_{\infty}}$ (see Eq.~\ref{flux_gauge_low}). We note that the quantity $\Gamma$ does not have an analytical expression in terms of the system parameters, but it is not independent from $g_1^{(f)}$, as both quantities grow with, e.g. the value of $x$. For sufficiently small $x$, the two quantities can be approximately linked as $g_1^{(f)} \approx  \Delta (\Gamma/f_{\rm{eg}})^{1/2}$, which makes the maximal interaction matrix element to be approximately given by $\Delta \times (\Gamma/f_{\rm{eg}})^{1/2}\times (\Delta/\Gamma)$. Assuming the superstrong coupling regime parameters $(\Delta/\Gamma) \approx 1/10$ and $\Gamma/f_{\rm{eg}} \approx 1/10$, and setting $\Delta = 100~\textrm{MHz}$ we would get the maximal interaction matrix element in the MHz range. Same estimate comes out from the rigorous numerical analysis for the proposed device parameters in Table~\ref{tableApp}. Increasing the cavity length would reduce the maximal interaction matrix element, but the number of coupled states will grow.

\section{Fine spectral structure example}
\label{section:fine}

In this section, we calculate the experimental spectroscopy signal (see Appendix~\ref{reflection_appendix} for detailed expressions), the main deliverable of our theory. We consider a one-port reflection measurement, which consists of sending a low-power coherent drive onto the cavity's port mirror (see Fig.~\ref{circuit_model}) and measuring the outgoing signal at the same frequency. We assume that the excited mode has an external quality factor due to the coupling to the measurement port and an internal quality factor due to the dielectric loss. The reflection magnitude $|S_{11}|$ can be expressed via these quality factors as well as the eigenvalues and eigenvector of the effective photon-photon interaction Hamiltonian~\ref{effective_Ham}. For simplicity, we ignore the lifetime of the two-particle excitations, assuming they are damped only due to their coupling to the one-paerticle excitation. We also ignore the decoherence of the fluxonium qubit due to the 1/f flux noise, which should be well under 100 kHz owing to its large loop inductance.

We show an example spectrum around the mode $k = 49$, in the single-particle approximation (Fig.~\ref{S11}a) and in the two-particle approximation ((Fig.~\ref{S11}b)). The two predictions are strikingly different by the fine-structure, which arises as a result of the hybridization of the single-particle excitation $p_{49}^{\dagger}|0\rangle$ with the following sequence of two-particle states: $p_{48}^{\dagger}b_{1}^{\dagger}|0\rangle$, $p_{47}^{\dagger}b_{2}^{\dagger}|0\rangle$, ..., $p_{34}^{\dagger}b_{15}^{\dagger}|0\rangle$. The matrix element $A_{49, 34}$ is practically zero, so there was no need to consider the sequence further, which justifies the separation of single-particle states into the dressed (high-frequency) and undressed (low-frequency) photons. The splittings reflect the interaction matrix elements in Eq.~\ref{effective_Ham}, while the visibility of the resonances is due to the single-particle component. As a general rule, a small interaction matrix element and a large detuning from the single-particle frequency results in a dimmer signal. 

The key condition for observing the predicted fine spectral structure in an experiment is that the two-particle splittings must exceed or at least become comparable to the natural linewidth of the cavity mode. For the mode in question at a frequency of about $8.7$ GHz we used the external quality factor $Q_{\textrm{ext}} = 2000$ and internal quality factor  $Q_{\textrm{int}} = 10000$, which makes the natural linewidth to be around 5 MHz. Such a relatively large loss rate is still plenty to resolve the two-particle hybridization. The loss rate in superconducting resonators can be easily made one or two orders of magnitude smaller. 

In principle, the described fine spectral structure in Fig.~\ref{S11}b can become even finer if we take into account the coupling to states with more than two particles. For example, one can consider a coherent cascade of first splitting a single photon into a superposition of two-photon pairs, in which the higher-frequency photons undergo the same splitting process again. Calculation wise, one would need to find the spectrum of the Hamiltonian~\ref{effective_Ham} by taking into account the three-particle states which share one low-frequency photon with the hybridized two-particle states. 

However, the hybridization with the three-particle states happens to be negligible due to a peculiarity of our example device. Namely, the bare modes frequencies are given by $\Delta (j - 1/2), j=1,2,...$ due to the ``open" type boundary condition chosen for the coupling mirror. Consequently, there is a kind of an even-odd effect for the energies of multi-particle states: the two-particle and the three-particle states bunch around different frequencies, separated by about $\Delta/2$. Consequently, those two-particle states that hybridized with the single-particle resonance $k=49$ are far off-resonant from any eligible three-particle states, which makes the fine spectral structure maximally simple and limited to only two-particle splittings.

\section{Conclusion and perspectives}
\label{section:conclusions}

We demonstrated that, surprisingly, the rotating-wave approximation can be broken already in the superstrong coupling regime of multi-mode cavity/circuit QED, conventionally defined by $\Delta \ll \Gamma \ll f_{\rm{eg}}$. This breakdown results in a dramatic fine spectral structure of multi-particle resonances in the cavity's linear (vanishing probe power) spectrum, in the $\Gamma$-vicinity of the qubit frequency $f_{\rm{eg}}$. It is essential that the qubit has a strong anharmonicity and a static dipole moment. The latter property breaks down the excitation parity symmetry and enables a hybridization between one-particle and two-particle states. The cavity's low-frequency modes are decoupled from the qubit in the single-particle approximation, but they are necessary to create a large number of nearly degenerate multi-particle excitations around the qubit resonance. While our 
prediction was worked out in detail for a specific case of a superconducting fluxonium qubit coupled to a transmission line resonator, the same effect should be relevant for any other cavity QED platform capable of superstrong coupling. 

The proposed photon-photon interaction model yields no direct hybridization between one-particle and three-particle states. It occurs indirectly as a second order process via the two-particle states. To check against the direct hybridization to three-particle states, we considered a scaled down version of our cavity system having only 6 modes. Fig.~\ref{benchmark} in Appendix ~\ref{section:benchmark} shows the comparison of the three-photon spectra obtained using our effective Hamiltonian~\ref{effective_Ham} and a numerical diagonalization of the ``microscopic" light-matter coupling Hamiltonian~\ref{H_tot} in both the charge and flux gauges. The two spectra show a negligible difference between the two-particle and three-particle splitting features. A direct three-particle conversion would likely become important as the ratio $\Gamma/f_{\rm{eg}}$ grows, in which case the system would enter the conventional ultrastrong coupling regime \cite{Koshino2022}. Calculating the fine spectral structure in that case remains a difficult open question.

The proposed circuit QED setup based on high-impedance transmission line and a fluxonium qubit potentially combines MHz-range conversion rates with 10-100 microsecond decoherence or absorbtion time (limited by the qubit and the cavity quality factors). To our knowledge such a parameter combination have not been possible so far in any system. 
The resulting ``strong down-conversion" makes the predicted spectral fine structure energy resolved and experimentally relevant. One application of this effect is a stringent experimental test for light-matter coupling models, which may help to resolve the debates around the proper gauge choice for the cavity QED beyond the strong coupling. Another general application of strong down-conversion lies in the realm of many-body physics. For example, one may wonder if the down-conversion cascade started by a single photon continues forever, that is the single photon dynamics involves an exponentially large number of multi-particle states, or if it gets stuck, say, at the two-photon states, as it was the case in the example presented here. This general question belongs to the subject of many-body localization~\cite{Altschuler1997}. Promising first experimental results in this direction have been recently reported~\cite{Kuzmin_2021, Nitish_Nature}.


\acknowledgements
We thank Erwan Roverc'h for discussions. V.E.M acknowledges support from US DOE Early Career
Award and from the ARO-MURI "Exotic States of Light in Superconducting Circuits" program. C.C. acknowledges financial support from FET FLAGSHIP
Project PhoQuS (grant agreement ID no.820392) and
from the French agency ANR through the project
NOMOS (ANR-18-CE24-0026), and TRIANGLE (ANR20-CE47-0011).

\appendix

\section{Foster's decomposition}
\label{foster_appendix}
Foster's theorem \cite{foster_1924} states that any physical admittance function $Y(\omega)$ can be represented by an infinite number of series $L_iC_i$-circuits connected in parallel, as shown in Fig. \ref{circuit_model}. The admittance  $Y^{F}(\omega)$ of such lump-element circuit is given by: 

\begin{equation}
    Y^{F}(\omega) = \sum_{i=1}^{\infty}  \frac{\mathrm{i} \omega C_i}{(1 - \omega^2 L_i C_i)}\,.
    \label{lumped}
\end{equation}
To find the values of $L_i$ and $C_i$, it is sufficient to match the poles and the residues of the two analytical functions $Y(\omega)$ and $Y^{F} (\omega)$.

\subsection{Open-ended non-dispersive line}

The admittance of an open section of a 1D transmission line of length $\ell$, characterized by the speed of light $v$ and the wave impedance $Z_{\infty}$ is given by \cite{Pozar}: 

\begin{equation}
    Y (\omega) = \frac{\mathrm{i}}{Z_{\infty}} \tan( \omega \ell /v)\,.
    \label{line}
\end{equation}
Matching the poles and residues of the functions given by Eq. (\ref{line}) and Eq. (\ref{lumped}) results in the equally spaced poles at frequencies $\omega_i = 2\pi \Delta (i-1/2)$, where the free spectral range is $\Delta = v/2l$. The inductances and capacitances read:

\begin{equation}
    L_i = L_1 =\frac{Z_{\infty}}{4\Delta} \, ,
\end{equation}

\begin{equation}
    C_{i} = \frac{1}{Z_{\infty} \pi^2 \Delta} \times \frac{1}{ (i - 1/2)^2}   \,.
\end{equation}

\subsection{Terminated dispersive line}

We can generalize the Foster's decomposition method for a line with an arbitrary wave dispersion $k(\omega)$, frequency-dependent wave impedance $Z_{\infty}(\omega)$ and an arbitrary reactive termination, characterized by the phase shift $\tau (\omega)$. The admittance of such a line is given by: 

\begin{equation}
    Y(\omega) = \frac{1}{Z_{\infty}(\omega)} \Big( \frac{\mathrm{e}^{-\mathrm{i} \tau(\omega) } -  \mathrm{e}^{-2\mathrm{i} k(\omega)\ell}}{\mathrm{e}^{-\mathrm{i} \tau(\omega) } +  \mathrm{e}^{-2\mathrm{i} k(\omega)\ell}}  \Big) \,,
    \label{admittance_dispersive}
\end{equation}
whose poles $\omega_j$ are implicitly defined by: 
\begin{equation}
     k(\omega_i)\ell = \pi (i-1/2) + \tau(\omega_i) \,.
    \label{phase_matching}
\end{equation}
In general, Eq. (\ref{phase_matching}) produces non-uniformly spaced modes. Nevertheless, the parameters $L_i$ and $C_i$ can still be expressed in a compact form:

\begin{equation}
    L_i = \frac{Z_{\infty}  (\omega_i)}{2} \frac{\ell \partial k}{\partial\omega}\bigg|_{\omega=\omega_i} \, ,
    \label{induct_arbit}
\end{equation}
\begin{equation}
    C_i = \frac{2}{ \omega^2_i Z_{\infty}(\omega_i)} \frac{\partial \omega}{\ell\partial k} \Big|_{\omega = \omega_i} \,.
    \label{cap_arbit}
\end{equation}

\subsection{Josephson transmission line example}
For the experimentally relevant case of a high-impedance line constructed from Josephson junction arrays, we get \cite{Kuzmin2019_1}:

\begin{equation}
    k(\omega)\ell = \frac{\omega /2\Delta}{\sqrt{(1 - (\omega / \omega_p)^2}} \, ,
\end{equation}x

\begin{equation}
    Z_{\infty} (\omega) = \frac{Z_{\infty}(\omega = 0)}{\sqrt{1 - (\omega / \omega_p)^2}} \, , 
\end{equation}
where $\Delta = (\partial \omega/\partial k)|_{\omega =0}/2\ell$ is the mode spacing in the limit $\omega \rightarrow 0$ and $\tau(\omega) = 0$, while $\omega_p$ is the plasma frequency of the chain junctions, which defines the high frequency propagation cutoff. We thus obtain the expressions

\begin{equation}
    L_i = \frac{Z_{\infty}(\omega = 0)}{4\Delta} \frac{1}{(1 - (\omega_i / \omega_p)^2)^2} \, ,
    \label{induc_JJline}
\end{equation}

\begin{equation}
    C_i = \frac{4\Delta }{\omega^2_i Z_{\infty}(\omega = 0) }(1 - (\omega_i / \omega_p)^2)^2 \, ,
    \label{cap_JJline}
\end{equation}
where the effect of the termination induced phase shift is taken into account in the values of $\omega_i$.

\section{Circuit quantization}
\label{section:quantization}
\begin{figure}[t!]
    \centering
    \includegraphics[scale=0.5]{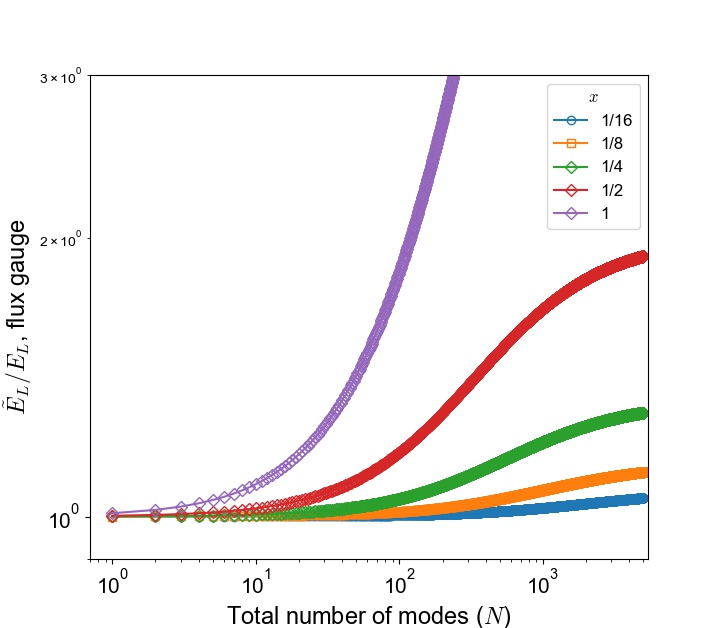}
    \caption{Relative renormalization of the inductive energy of the fluxonium atom in the flux gauge as a function of the total number of modes ($N$) included in the Foster's decomposition of the transmission line. Same parameters as in Fig. \ref{couplings}. The legend indicates the values of the geometric fraction $x$.
    \label{flux_renorm}}
\end{figure}
\label{hybrid_appendix}

Here we derive the circuit Hamiltonian for our multi-mode cavity QED system in the mixed gauge, whose variable choice is shown in Fig. ~\ref{circuit_model} (d). Note that the choice of the parameter $i_{0}$ interpolates between the pure charge gauge $(i_{0} = 0)$ and the pure flux gauge $(i_{0} = N)$. The variable choices for these specific cases are shown in Fig. ~\ref{circuit_model} (b, c).

\subsection{Derivation of the circuit Hamiltonian}
We link the dependent variable $\varphi_M$ to the independent variables $\varphi_i$ with $i = 1, 2, \dots N$. Using the current conservation at node $M$, marked by the gray circle in Fig. ~\ref{circuit_model} (d), we get:

\begin{equation}
    \frac{\varphi_M}{{x}L} + \frac{\varphi_M}{(1 - {x})L} + \sum_{i=1}^{i_{0}} \frac{\varphi_M}{L_i} = \frac{\varphi_J}{(1 - {x})L} + \sum_{i=1}^{i_{0}} \frac{\varphi_i}{L_i} - \sum_{i > N} \frac{\varphi_i}{L_i} \, .
    \label{Kirchoff_hybrid}
\end{equation}
We can now write the circuit's classical Lagrangian $\mathfrak{L} = T - V$ in terms of the generalized fluxes $\phi_i = \varphi_i(\hbar/2e)$ and $\phi_J = \varphi_J(\hbar/2e)$.
The kinetic and potential energy contributions to the Lagrangian are given by:
\begin{equation}
    T = \frac{1}{2}\boldsymbol{\dot{\phi}}^{T} \mathcal{C}   \boldsymbol{\dot{\phi}} \, ,
\end{equation}

\begin{equation}
    V = \frac{1}{2} \boldsymbol{\phi}^{T} \mathcal{L}^{-1} \boldsymbol{\phi} - E_J \cos(\varphi_J - \varphi_{\textrm{ext}}) \,,
\end{equation}

\noindent where we have defined the vector 
\begin{equation}
\boldsymbol{\phi} \equiv (\phi_0 = \phi_J, \phi_1, \phi_2, \dots \phi_N) \, .
\end{equation}
The symmetric capacitance matrix $\mathcal{C}$ and the symmetric inductance matrix $\mathcal{L}$ have the following matrix elements:

\begin{equation}
    {{\mathcal C}_{0,0}} = C_J + {\tilde{x}}^2  \sum_{j > i_{0}} C_j
    \, ,
\end{equation}
\begin{equation}
    {{\mathcal C}_{0,0 < i \leq i_{0} }} =  - \tilde{x} \frac{L_{\sum}}{L_i} \sum_{j > i_{0}} C_j \, ,
\end{equation}
\begin{equation}
    {{\mathcal C}_{0,i> i_{0} }} = -\tilde{x} C_i - \tilde{x} \frac{L_{\sum}}{L_i} \sum_{j > i_{0}} C_j \, ,
\end{equation}

\begin{gather}
    {{\mathcal C}_{i \neq 0,j \neq 0}} = C_i \delta_{i,j} +  \frac{L_{\sum}}{L_j} C_i \Theta[i - i_{0} -1] \\
    + \frac{L_{\sum}}{L_i} C_j \Theta[j - i_{0} -1] \nonumber 
    + \frac{L_{\sum}}{L_i} \frac{L_{\sum}}{L_j} \sum_{k > i_{0}} C_k \, ,
\end{gather}

\begin{equation}
    (\mathcal{L}^{-1})_{0,0} = \frac{(1- \tilde{x})}{(1-x)L} ,
    \label{renormalized_L}
\end{equation}
\begin{equation}
    (\mathcal{L}^{-1})_{0,0 < i \leq i_{0}} = \frac{\tilde{x}}{L_i} \, ,
\end{equation}

\begin{equation}
    (\mathcal{L}^{-1})_{0,i >  i_{0}} = 0  \, ,
\end{equation}

\begin{gather}
    (\mathcal{L}^{-1})_{i \neq 0,j \neq 0} = \delta_{i, j} \frac{1}{L_i} \\
    \nonumber + \frac{L_{\sum}}{L_i L_j} \Theta[i - i_{0} -1] \Theta[j - i_{0} -1] \,,
\end{gather}
where the variables $L_{\sum}$ and $\tilde{x}$ are given by:
\begin{equation}
    L_{\sum} = \Bigg( \frac{1}{xL}  + \frac{1}{(1-x)L} + \sum_{i=1}^{i_{0}} \frac{1}{L_i}  \Bigg)^{-1} \, , 
\end{equation}

\begin{equation}
    \tilde{x}= \frac{L_{\sum}}{(1-x)L}.
\end{equation}
 
\noindent Here $\Theta$ is the discrete Heaviside step function. Note that for the case of pure charge gauge, $L_{\sum} = x(1-x) L$ and $\tilde{x} \to x$.

In order to derive the Hamiltonian of this circuit, we define the conjugate momenta (charges) to the generalized coordinates shown in Fig. ~\ref{circuit_model}:
\begin{equation}
    \boldsymbol{q} = \frac{\partial \mathfrak{L}}{\partial \boldsymbol{\dot{\phi}}} = {\mathcal C} \boldsymbol{\dot{\phi}}\, .
\end{equation}

The Hamiltonian can be obtained by applying a Legendre transform to the Lagrangian: 
\begin{equation}
    H = \boldsymbol{q}^T \boldsymbol{\dot{\phi}} - \mathfrak{L} = 
     \frac{1}{2} \boldsymbol{q}^T {\mathcal C}^{-1} \boldsymbol{q} + \frac{1}{2} \boldsymbol{\phi}^{T} \mathcal{L}^{-1} \boldsymbol{\phi} - E_J \cos(\varphi_J - \varphi_{\textrm{ext}})  \, .
\end{equation}
The circuit Hamiltonian can be written as a sum of three parts. The atomic part of the Hamiltonian reads: 
\begin{equation}
    H_{\textrm{atom}} = 4E_C n^2_J + \frac{1}{2} \tilde{E}_L \varphi^2_J - E_J \cos(\varphi_J - \varphi_{\textrm{ext}}) \, ,
    \label{H_atom_app}
\end{equation}
where $E_C = e^2 (\mathcal{C}^{-1} )_{0,0}/2$ and $\tilde{E}_L = (\hbar/2e)^2 (\mathcal{L}^{-1})_{0,0}$, which formally depend on the $(0,0)$ element of the inverse capacitance and inductance matrices. Remarkably, it can be proved analytically that  $(\mathcal{C}^{-1})_{0,0} = 1/C_J$, irrespective of choice of $i_{0}$ and  $(\mathcal{L}^{-1})_{0,0} = 1/L$ for $i_{0} = 0$.
This is relevant, because it shows that the atom Hamiltonian defined in Eq. (\ref{H_atom}) is not renormalized by the transmission line parameters and depends only on the bare fluxonium parameters for the pure charge gauge. Note that in the mixed gauge the atom inductive energy is renormalized and is a function of both $x$ and $i_{0}$. The worst case scenario occurs for the flux gauge where the inductive energy of the atom now depends on the number of modes included in the lumped element Foster's decomposition of the transmission line. The renormalization of the inductive energy as a function of the total number of modes included in the flux gauge is shown in Fig.~\ref{flux_renorm}.

The photonic part of the  Hamiltonian, which is quadratic in the charge and flux variables, reads:

\begin{equation}
    H_{\textrm{modes}} = \frac{1}{2}  \sum_{i,j=1}^{N} {({\mathcal C} ^{-1})_{i,j} } \, q_i q_j + \frac{1}{2}  \sum_{i,j=1}^{N} ({\mathcal L}^{-1})_{i,j} \phi_i \phi_j\,. 
\end{equation}

Finally, the interaction between the fluxonium degrees of freedom and the transmission line modes is given by:
\begin{equation}
    H_{\textrm{int}} = q_J\sum_{i=1}^{N} ({\mathcal C}^{-1})_{0, i}  q_i-\phi_J\sum_{i=1}^{N} ({\mathcal L}^{-1})_{0, i}  \phi_i \,.
\end{equation}

\subsection{Renormalization due to the diamagnetic terms}

In order to quantize the classical Hamiltonian we introduce photon annihilation (creation) operators $\hat{a}_i (\hat{a}^{\dagger}_i)$ as: 
\begin{equation}
    \hat{q}_i = {\mathrm i} \sqrt{\frac{\hbar}{2 R_Q}} (\hat{a}_i - \hat{a}^{\dagger}_i)\, ,    
\end{equation}

\begin{equation}
    \hat{\phi}_i = \sqrt{\frac{\hbar R_Q}{2} } (\hat{a}_i + \hat{a}^{\dagger}_i)\, ,
\end{equation}
where $R_Q = h/4e^2$ is the superconducting resistance quantum. Note that any other resistance value could be used, as the dummy operators $\hat{a}_i$ ($\hat{a}_i^{\dagger}$) merely define a starting basis for the cavity modes. 
The operator $\hat{H}_{\textrm{modes}}$ 
 can be diagonalized via a Bogoliubov transformation, because it is a quadratic function of the photon operators. To proceed,  
we introduce a vector of annihilation operators $\boldsymbol{\hat{a}}$:
\begin{equation}
        \boldsymbol{\hat{a}}^{T} \equiv (\hat{a}_1, \hat{a}_2 \dots \hat{a}_N) \,.
\end{equation}
The quantized Hamiltonian $\hat{H}_{\textrm{modes}}$ can be rewritten in the following matrix form:
\begin{equation}
\hat{H}_{\textrm{\rm modes}} = \frac{1}{2} 
(\boldsymbol{\hat{a}}^{\dagger, T}, \boldsymbol{\hat{a}}^T) {\mathcal \eta} {\mathcal M} \begin{pmatrix} \boldsymbol{\hat{a}} \\
\boldsymbol{\hat{a}}^{\dagger}
\end{pmatrix}
 = \frac{1}{2} 
(\boldsymbol{\hat{b}}^{\dagger, T}, \boldsymbol{\hat{b}}^T) {\mathcal \eta} \Omega \begin{pmatrix} \boldsymbol{\hat{b}} \\
\boldsymbol{\hat{b}}^{\dagger}
\end{pmatrix}
\, ,
\end{equation}
where ${\mathcal \eta}$ is the diagonal $2N \times 2N$ matrix such that ${\mathcal \eta}_{ii} = 1$ for $i \leq N$ and ${\mathcal \eta}_{ii} = -1$ for $i > N$. The other $2N \times 2N$ matrix to deal with is:
\begin{equation}
\label{bog}
    {\mathcal M} = \begin{pmatrix} {\mathcal X} & {\mathcal Y} \\
   - {\mathcal Y} & - {\mathcal X} \end{pmatrix}   \, .
\end{equation}
The symmetric $N \times N$ matrices ${\mathcal X}$ and ${\mathcal Y}$ are defined as:
 
\begin{equation}
    {\mathcal X}_{i, j} =R_Q ({\mathcal L}^{-1})_{i,j} + R^{-1}_Q ({\mathcal C}^{-1})_{i,j} \,,
\end{equation}

\begin{equation}
    {\mathcal Y}_{i, j} = R_Q ({\mathcal L}^{-1})_{i,j} - R^{-1}_Q ({\mathcal C}^{-1})_{i,j} \, ,
\end{equation}
where $i, j \in \{1,2, ..., N\}$.
Finally the diagonal matrix $\Omega = diag(\omega_1,\omega_2, .., \omega_N, -\omega_1,-\omega_2, ..., -\omega_N)$ contains the frequencies of the normal Bogoliubov modes. It is defined as ${\mathcal M} = \mathcal{Q} \Omega \mathcal{Q}^{-1}$, so that the Bogoliubov mode operators are:

\begin{equation}
\begin{pmatrix} \boldsymbol{\hat{b}} \\
\boldsymbol{\hat{b}}^{\dagger}
\end{pmatrix} = 
\mathcal{Q}^{-1} 
\begin{pmatrix} \boldsymbol{\hat{a}} \\
\boldsymbol{\hat{a}}^{\dagger} \,
\end{pmatrix} \, ,
\end{equation}
where the matrix $\mathcal{Q}$ can be expressed in the form   
\begin{equation}
    {\mathcal Q} = 
    \begin{pmatrix} {V} & {U} \\
   - {U} & - {V} 
   \end{pmatrix}   \, .
\end{equation}

The Hamiltonian $\hat{H}_{\textrm{modes}}$ can be diagonalized in terms of bosonic normal-mode operators $\hat{b}_i$ ($\hat{b}_i^{\dagger}$):
\begin{equation}
\hat{H}_{\textrm{modes}} =  \sum_{i = 1}^N
\hbar \omega^{(m)}_{i} \hat{b}^{\dagger}_{i}  \hat{b}_{i} \, ,
\label{H_modes}
\end{equation}

\noindent where $\omega^{(m)}_i$ are the corresponding normal-mode frequencies.

The coupling of the atom to the transmission line is represented by the interaction Hamiltonian:
\begin{gather}
    \hat{H}_{\textrm{int}} ={\mathrm{i}}\hat{n}_J  \frac{4}{\pi^{1/2}} E_C\sum_{i=1}^{N}    ({\mathcal C}^{-1} C_J)_{0,i}  (\hat{a}_i - \hat{a}^{\dagger}_i) \\ \nonumber
    -\hat{\varphi}_J\frac{E_{L}}{4 \pi^{3/2}} \sum_{i=1}^{N} (\mathcal{L}^{-1} L)_{0,i} (\hat{a}_i + \hat{a}^{\dagger}_i)
    \, .
\end{gather}
In order to express $\hat{H}_{\textrm{int}}$ in terms of the normal-mode transmission line operators $\hat{b}_{\nu}$, we need to invert the Bogoliubov transformation as follows:
\begin{equation}
    \hat{a}_i = \sum_{\nu=1}^{N} V_{i,\nu} \hat{b}_{\nu} + \sum_{\nu=1}^{N} U_{i, \nu} \hat{b}^{\dagger}_{\nu} \, .
\end{equation}

Note that the conservation of the bosonic commutation rules, i.e., $[{b}_{\nu},{b}_{\nu'}^{\dagger} ] = \delta_{\nu,\nu'}$, imposes the following normalization condition for the Bogoliubov transformation coefficients:
\begin{equation}
    \sum_{i=1}^{N} U_{i, \nu}^2 - V_{i, \nu}^2 = 1 \, .
\end{equation}

\subsection{The mixed gauge QED Hamiltonian}

We have now all the ingredients to formulate the multi-mode cavity QED Hamiltonian in terms of the fluxonium  operators $\hat{\varphi}_J$, $\hat{n}_J$, and the transmission line normal mode operators $\hat{b}_i \: (\hat{b}^{\dagger}_i)$ (that include already diamagnetic-like renormalizations). The total Hamiltonian is $\hat{H} = \hat{H}_{
\textrm{atom}}+ \hat{H}_{\textrm{modes}}+\hat{H}_{\textrm{int}}$, where $\hat{H}_{\textrm{atom}}$ is given by Eq.  (\ref{H_atom_app}), $\hat{H}_{\textrm{modes}}$ is given by Eq. (\Ref{H_modes}) and $\hat{H}_{\textrm{int}}$ can be expressed as

\begin{gather}
     \hat{H}_{\textrm{int}} ={\mathrm{i}} \hat{n}_J  \sum_{i=1}^{N} h g^{(m, c)}_i   (\hat{b}_i - \hat{b}^{\dagger}_i)  \\
     \nonumber
     -\varphi_J \sum_{i=1}^{N} h g^{(m,f)}_i (\hat{b}_i + \hat{b}^{\dagger}_i)
     \,, \label{H_{int}}
\end{gather}

with the mixed gauge coupling constants given by:
\begin{equation}
    h g^{(m,c)}_i = \frac{4}{\pi^{1/2}}E_C \sum_{j=1}^{N} ({\mathcal C}^{-1}C_J)_{0,j} (V_{j,i} - U_{j,i})    \,, 
     \label{couplings_charge}
\end{equation}
\begin{equation}
    h g^{(m,f)}_i = \frac{1}{4 \pi^{3/2}}E_{L} \sum_{j=1}^{N} ({\mathcal L}^{-1} L)_{0,j} (V_{j,i} + U_{j,i})    \,. 
     \label{couplings_flux}
\end{equation}

Note that for the special case of $i_{0} = 0$ (charge gauge) the flux coupling constants are identically zero and for the case of $i_{0} = N$ (flux gauge) the charge coupling constants are identically zero. Also note that the flux coupling constants $g^{(m,f)}_i$ can be written analytically for $i< i_{0}$:

\begin{equation}
    hg^{(m,f)}_{i<i_0} = \hbar \omega^{(m)}_i x \sqrt{\frac{R_Q}{4 \pi Z_{\infty} (i - 1/2)}}
    \label{flux_gauge_low} \, .
\end{equation}

\section{Effective Hamiltonian}
\label{section:analytical}
Here we provide the detailed derivation of the effective photon-photon Hamiltonian mediated by the fluxonium atom. In doing so, we only consider the transition from the ground state $\vert {\rm g} \rangle$ to the first excited state $\vert {\rm e} \rangle$. Within such two-level approximation, the fluxonium atom Hamiltonian can be approximated as:
\begin{equation}
    \hat{H}_{\rm{atom}} = \hbar \omega_{\rm{eg}} (\varphi_{\rm{ext}}) \vert \rm{e} \rangle \langle \rm{e} \vert .
    \label{H_atom}
\end{equation}

The flux-dependent transition frequency $\omega_{\rm{eg}}$ in Eq. (\ref{H_atom}) is calculated by a numerical diagonalization of the full fluxonium Hamiltonian given in Eq. (\ref{fluxonium}). We divide the transmission modes of the system into two sets. The first set includes modes with frequencies much smaller than the fluxonium qubit transition frequency at $\varphi_{\rm{ext}} = \pi$ (mode index $i$ such that $1 \leq i \leq i_{0}$). The second set of modes includes those that quasi-resonantly hybridize with the qubit transition. We will denote $k$ the index of such modes where $ i_{0} < k \leq N$. To write the complete Hamiltonian of the system, we choose the mixed gauge described in section \ref{section:system} (C) with the separation index being $i_{0}$. In a first step, we ignore for the time being the coupling of the fluxonium to the low-frequency bosonic modes of the transmission line and diagonalize the following Hamiltonian
\begin{widetext}
\begin{equation}
    \label{H_0}
    \hat{H}_0 = \hbar \omega_{\mathrm{eg}} \vert \mathrm{e} \rangle \langle \mathrm{e} \vert + \sum_{i =1}^{N}\hbar \omega^{(m)}_i \hat{b}^{\dagger}_i \hat{b}_i - \hat{\varphi}_J \sum_{i > i_{0}} h g^{(m,f)}_i (\hat{b}_i + \hat{b}^{\dagger}_i) + \mathrm{i} \hat{n}_J \sum_{i > i_{0}} h g^{(m,c)}_i (\hat{b}_i - \hat{b}^{\dagger}_i) \,,
\end{equation}
\end{widetext}
in the single-excitation subspace spanned by the basis states:
\begin{equation}
{\mathcal B }_0 = \left \{ \vert \mathrm{e} \rangle \vert 0 \rangle, \vert \mathrm{g} \rangle  \hat{b}^{\dagger}_{k} \vert 0 \rangle \right \}_{1 < k \leq N}  \, .
\label{B0}
\end{equation}
Note that the single-excitation eigenstates of the Hamiltonian (\ref{H_0}) consist of bare transmission line modes for $k \leq i_{0}$ with excitation frequency $\omega^{(m)}_k / 2\pi$. For $k > i_{0}$ they have the form:
\begin{equation}
\hat{p}_k^{\dagger} \vert G \rangle = 
W_{k,0} \vert \mathrm{e} \rangle \vert 0 \rangle + 
\sum_{k'} W_{k,k'} \vert \mathrm{g} \rangle  \hat{b}^{\dagger}_{k'} \vert 0 \rangle \, ,
\end{equation}
where $\Omega_k / 2 \pi$ is the corresponding single-polariton excitation frequency. We use the term ``polariton" state in analogy to similar states appearing in semiconductor systems. The Hamiltonian in Eq. (\ref{H_0}) can now be rewritten as 

\begin{equation}
    \label{H_0_d}
    \hat{H}_0 = \sum_{i=1}^{i_{0}} \hbar \omega^{(m)}_i \hat{b}_i^{\dagger} \hat{b}_i + \sum_{k > i_{0}} \hbar \Omega_k \hat{p}^{\dagger}_k \hat{p}_k \,.
\end{equation}

As a second step of our procedure, we now include the effect of the coupling of low-frequency modes to the fluxonium in Eq. (\ref{H_0_d}). The coupling term is given by 

\begin{equation}
    \hat{V} =  - \hat{\varphi}_J \sum_{i = 1}^{i_{0}} h g^{(m,f)}_i (\hat{b}_i + \hat{b}^{\dagger}_i) + \mathrm{i} \hat{n}_J \sum_{i = 1}^{i_{0}} h g^{(m,c)}_i (\hat{b}_i - \hat{b}^{\dagger}_i) \,.
\end{equation}
As it can be directly inspected, this term gives rise to an effective coupling between the bare transmission line modes and the polaritonic modes via a three-wave mixing process. Indeed, this elementary process converts one excitation into two excitations, one in an another polaritonic mode and one in a low-frequency photon mode. The corresponding coupling is quantified by the following matrix element:
\begin{equation}
    \langle G \vert \hat{p}_k\hat{V}\hat{p}^{\dagger}_{k'} \hat{b}^{\dagger}_i \vert G \rangle = h g^{(m, f)}_{i} W^{*}_{k,0} W_{k', 0} (\langle \mathrm{e} \vert \hat{\varphi}_J \vert \mathrm{e} \rangle -  \langle \mathrm{g} \vert \hat{\varphi}_J \vert \mathrm{g} \rangle)  \,.
    \label{2-photon_matrix}
\end{equation}
This finally leads to the effective Hamiltonian of our multi-mode cQED system reported in Eq. (\ref{effective_Ham}) with the coupling constants $A_{kk'}$ and $g$ given respectively by (\ref{Akk}) and (\ref{g}).

\begin{figure}[t!]
    \centering
    \includegraphics[width =0.5\textwidth]{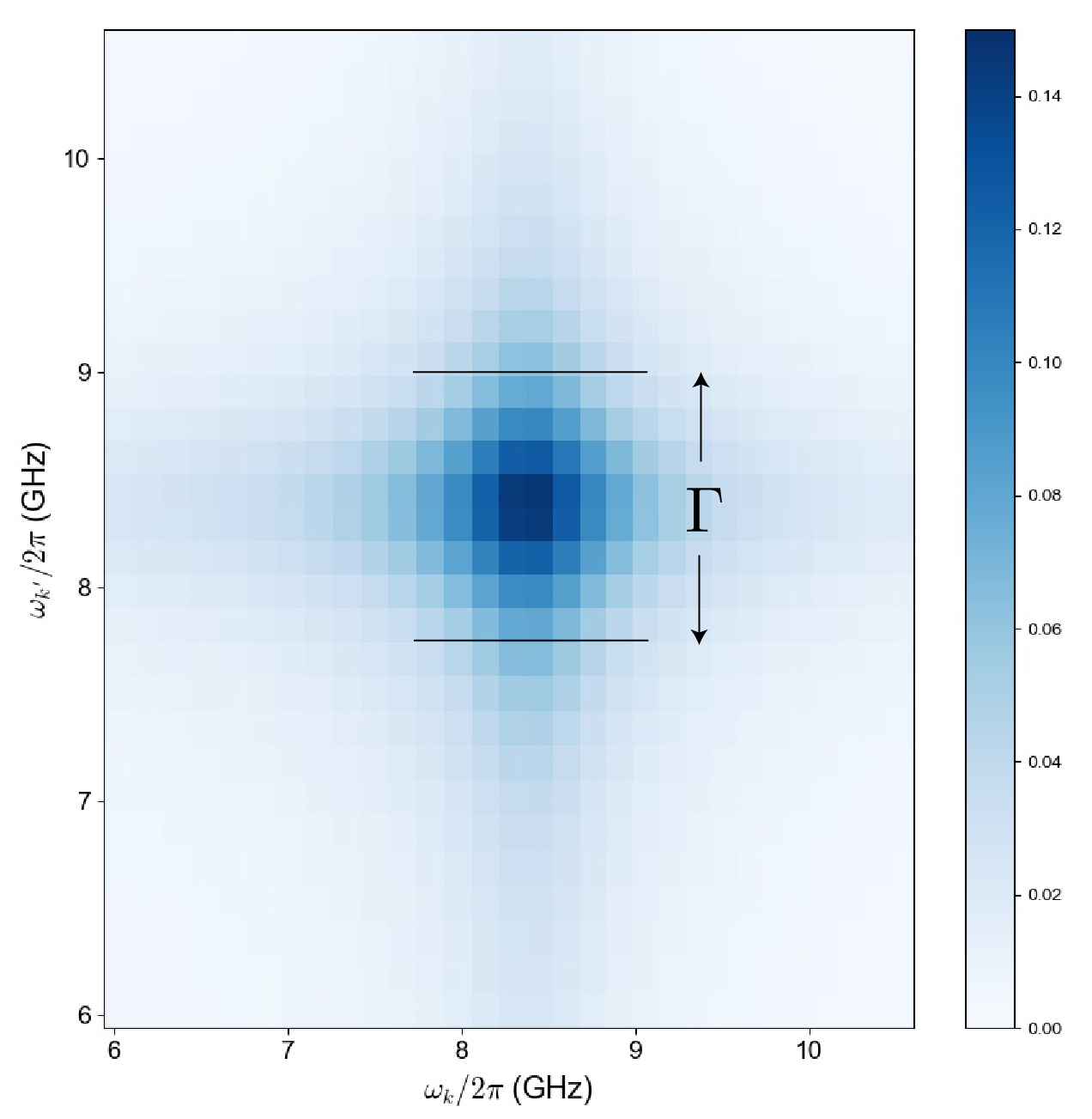}
    \caption{Color plot of $A_{k, k'}$, defined in Eq. (\ref{Akk}), the fluxonium component in polariton modes $k$ and $k'$  as a function of the corresponding polariton frequencies when the fluxonium $\vert \mathrm{e} \rangle \leftrightarrow \vert \mathrm{g} \rangle$ transition frequency $\omega_{\mathrm{eg}}/ 2 \pi$ is tuned to the value  $8.2 \: \mathrm{GHz}$. Same parameters as Fig. \ref{couplings} (a) and (b).
     }
    \label{qubit_comp}
\end{figure}

The three-wave-mixing strength $g$ is proportional to the coupling of the fluxonium to the low frequency mode $i$.  In the flux gauge, it is given by $g^{(m,f)}_i = g^{(m,f)}_1 \sqrt{i}$, as it can be seen from Eq. (\ref{flux_gauge_low}). In addition, the three-wave-mixing strength is also proportional to $\langle \mathrm{e} \vert \hat{\varphi}_J \vert \mathrm{e} \rangle -  \langle \mathrm{g} \vert \hat{\varphi}_J \vert \mathrm{g} \rangle)$, which is finite only when there is an asymmetric fluxonium potential. Moreover, the coupling constant $A_{kk'}$ depends on the fluxonium components of the polaritonic modes participating in the three-wave-mixing process (see Fig. ~\ref{qubit_comp}). 

The effective Hamiltonian can now be used to construct the full spectrum of our system by an iterative procedure that consists of enlarging the Hilbert space to include multi-photon states. Two-excitation states can be accounted for by considering the subpace spanned by the following basis of states:
\begin{equation}
{\mathcal B }_1 = {\mathcal B }_0 \bigcup 
\left \{ 
\hat{p}^{\dagger}_{k} \hat{b}^{\dagger}_{i} 
\vert 0 \rangle \right \}^{i_{0} < k \leq N}_{1 \leq i \leq i_{0}}   \, .
\label{B1}
\end{equation}

In this expanded Hilbert space, the single-polariton states are no longer eigenstates of the Hamiltonian. This is due to the fact that the single-polariton states now can hybridize with two-photon states due to the photon-photon interactions mediated by the fluxonium. Given the large number of frequency degeneracies between single-polariton states and two-photon states,  determined by the relation $\Omega_{k} = \Omega_{k'} + \omega_i$, the  diagonalization  of  the effective Hamiltonian produces a myriad of energy anticrossings in the spectrum. If we want to include the effect of three-photon states, we need to enlarge the basis of the Hilbert space by considering the states:

\begin{equation}
{\mathcal B }_2 = {\mathcal B }_0 \bigcup {\mathcal B }_1 \bigcup 
\left \{ 
\hat{p}^{\dagger}_{k} \hat{p}^{\dagger}_{k'} \hat{b}^{\dagger}_{i} 
\vert 0 \rangle \right \}^{i_{0} < k,k' \leq N}_{1 \leq i \leq i_{0}}   \, .
\label{B1}
\end{equation}

This process of expanding the Hilbert space can be iterated to four-particle states and so on so forth.

\begin{figure*}[t!]
    \centering
    \includegraphics[width=1\textwidth]{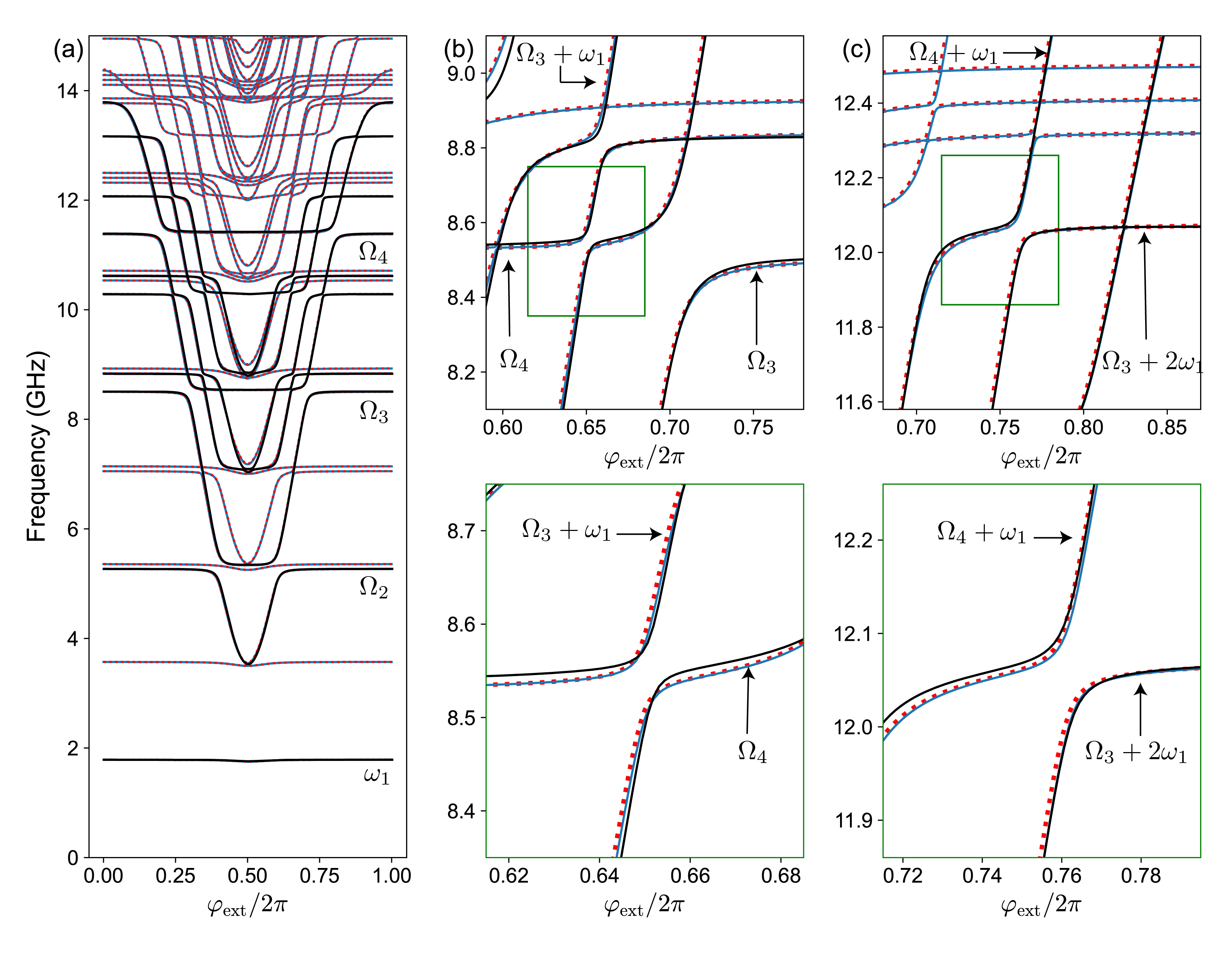}
    \caption{
    (a) Excitation spectrum of the multi-mode circuit QED system (parameters of cavity and fluxonium are in the text) for $N = 6$ unequally spaced modes obtained by numerical diagonalization in the charge (red dotted lines) and flux (solid blue lines) gauge. These results have been obtained by including the $7$ lowest-energy fluxonium eigenstates. The spectrum obtained with the effective Hamiltonian (Eq.~\ref{effective_Ham}) in the mixed gauge is displayed by the solid black lines. Note that the curves in the three gauges with the considered truncation parameters are all in very good agreement (deviations around a few percent). (b) Zoom in of the spectrum near the anticrossing of mode $k=3$ with the bare $\vert \rm{e} \rangle \to \vert \rm{g} \rangle$ transition, showing hybridization between the two-photon state with frequency $(\Omega_3  + \omega_1)$ and a single-polariton state with frequency $\Omega_4$. The red dotted line is the result in the charge gauge with $7$ levels, while the black solid line shows the energy levels calculated using the effective Hamiltonian described in Eq. (\ref{effective_Ham}) (mixed gauge, two-level approximation for the fluxonium). (c) Same as in (b), but this time showing the anti-crossing between the two-photon state with frequency $(\Omega_4 + \omega_1)$ and the three-photon state with frequency $(\Omega_3 + 2\omega_1)$. The parameters are the same as in Fig. \ref{couplings} (a) and (b), but with the transmission line length reduced by a factor $20$ and $x = 0.2$.}
    \label{benchmark}
\end{figure*}

\section{Exact diagonalization benchmark}
\label{section:benchmark}
Here we benchmark the effective Hamiltonian in Eq. (\ref{effective_Ham}) in the mixed gauge, whose derivation is detailed in Appendix \ref{section:analytical}, and the procedure to calculate the eigenspectrum of our multi-mode circuit QED system by comparison to brute-force diagonalization for finite-size systems. A full numerical diagonalization of the Hamiltonian in Eq. (\ref{H_tot}) rapidly becomes a daunting task, because the Hilbert space size grows exponentially with the number $N$ of bosonic modes and the number of photons involved. Here we present results for a system with $N = 6$ bosonic modes. Moreover, we also report  results in the other two gauges (charge and flux gauge) including multiple fluxonium levels and show that the correct results can be obtained by approximating the fluxonium as a two-level system in the mixed gauge.

In our numerical calculations, we consider a fluxonium atom and transmission line described by the parameters summarized in Table 1. In this example the line is 20 times shorter and shares a fraction $x = 0.2$ of the inductance with the fluxonium. The unequal spacing of the modes in the transmission line is achieved simply by introducing the ultra-violet cutoff for wave propagation due to the plasma frequency $\omega_p/2\pi = 25 \: \mathrm{GHz}$. The unequal mode spacing breaks the degeneracy of different multi-photon states and isolates anti-crossings between any two given states in the many-body spectrum of the system. 

Note that we have also derived the circuit Hamiltonian of the system in both the flux and the charge gauges following the procedure described in Appendix \ref{section:quantization}. We have diagonalized the circuit Hamiltonian in both gauges using the lowest $40$ levels of the uncoupled photonic modes of the transmission line and the $7$ lowest-energy levels of the fluxonium. In Fig. \ref{benchmark} (a), we report the energy eigenvalues in such two gauges with the prediction of the effective Hamiltonian (mixed gauge, two-level approximation for the fluxonium). The excellent agreement between the three gauges in Fig. \ref{benchmark} (a)) not only illustrates the gauge invariance of our multi-mode cQED model, but also serves as the additional proof that we have used a large enough Hilbert space for the diagonalization of the Hamiltonian. In particular, it also validates the two-level approximation for the fluxonium in the mixed gauge. A zoom in of the eigenspectrum  plot near the vacuum Rabi splitting of the third photonic mode reveals an additional anti-crossing (Fig. \ref{benchmark} (b)). This anti-crossing can be explained as the spectral manifestation of the hybridization between the single-polariton state $ \hat{p}^{\dagger}_{4} \vert G \rangle $ and the two-photon state with one excitation in polariton mode $k = 3$ and one excitation in mode $k = 1$, namely the state $ \hat{p}^{\dagger}_{3} \hat{b}^{\dagger}_1 \vert G \rangle $. To reproduce these spectral features with the effective Hamiltonian derived in Appendix \ref{section:analytical},  first we diagonalize the Hamiltonian only in the single-excitation subspace, where we ignore the coupling of the fluxonium to the lowest transmission line mode, and obtain the polaritonic frequencies $\Omega_k$ for $k > 1$. Then we include the effect of the coupling of the lowest mode by expanding the Hilbert space to the two-photon and three-photon states. We finally calculate the spectrum by diagonalizing the following $7 \times 7$ matrix:

\begin{widetext}
    \begin{equation}
    H_{\mathrm{eff}} = \begin{pmatrix}
       \Omega_3/2 \pi & 0 & 0 & g^{(m,f)}_1 A_{3, 3} & g^{(m,f)}_1 A_{3, 4} & 0 & 0  \\
       0 & \Omega_4 /2 \pi & 0  &  g^{(m,f)}_1 A_{4, 3} & g^{(m,f)}_1 A_{4, 4} & 0 & 0\\
       0 & 0 & \Omega_5/ 2 \pi & g^{(m,f)}_1 A_{5, 3} & g^{(m,f)}_1 A_{5, 4} & 0 & 0 \\
        g^{(m,f)}_1 A_{3, 3} & g^{(m,f)}_1 A_{3, 4} & g^{(m,f)}_1 A_{3, 5} & \Omega_3/ 2 \pi + \omega_1 / 2\pi & 0 & \sqrt{2}  g^{(m,f)}_1 A_{3, 2} & \sqrt{2} g^{(m,f)}_1 A_{3, 3}  \\
        g^{(m,f)}_1 A_{4, 3} & g^{(m,f)}_1 A_{4, 4} & g^{(m,f)}_1 A_{4, 5} & \Omega_4/ 2 \pi + \omega_1 / 2\pi & 0 & \sqrt{2}  g^{(m,f)}_1 A_{4, 2} & \sqrt{2} g^{(m,f)}_1 A_{4, 3} \\
        0 & 0 & 0 & \sqrt{2}  g^{(m,f)}_1 A_{2, 3} & \sqrt{2} g^{(m,f)}_1 A_{2, 4} & \Omega_2/2\pi + 2 \times \omega_1/2 \pi & 0 \\
        0 & 0 & 0 & \sqrt{2}  g^{(m,f)}_1 A_{3, 3} & \sqrt{2} g^{(m,f)}_1 A_{3, 4} & 0 &\Omega_3/2\pi + 2 \times \omega_1/2 \pi  \\
    \end{pmatrix} \\.
\end{equation}
\end{widetext}

The solid black line in Fig. \ref{benchmark} (b, c) shows the corresponding results, which are in excellent agreement with the full numerical diagonalization. 
Fig. \ref{benchmark} (c) highlights an anticrossing between the two-photon state $\hat{p }^{\dagger}_4 \hat{b}^{\dagger}_1 \vert G \rangle$ and three-photon state $\hat{p}^{\dagger}_3 (\hat{b}^{\dagger}_1)^2 \vert G \rangle$. 

\section{Calculation of reflection spectra}

\label{reflection_appendix}
To obtain the microwave reflection spectra, we first write the linear reflection coefficient for the bare transmission line modes: 

\begin{equation}
S_{11}^{\textrm{bare}} (\omega) =\prod_{i=1}^{N} \frac{2 \mathrm{i} (\omega - \omega_i)/ \omega_i - (Q_{i,{\mathrm{ext}}})^{-1} + (Q_{i,{\mathrm{int}}})^{-1} } {2 \mathrm{i} (\omega - \omega_i)/ \omega_i + (Q_{i,\mathrm{ext}})^{-1} + (Q_{i,\mathrm{int}})^{-1}}\, ,
\label{ref_coe_bare}
\end{equation} 
 where $Q_{i,\mathrm{int}}$ and $Q_{i,\mathrm{ext}}$ are respectively the internal and external quality factors of the bare transmission line modes. 
To calculate the reflection coefficient in the presence of the fluxonium atom, the expression is replaced by the product over the eigenstates of the total Hamiltonian in the Hilbert subspace $\mathcal{B}_s$, meaning that $(s+1)$-photon states are included. Namely, we have: 
\begin{equation}
S_{11}^{(s)}(\omega) =\prod_{k=1}^{N^{(s)}} \frac{2 \mathrm{i} (\omega - \tilde{\omega}_k)/ \tilde{\omega}_k - (\tilde{Q}_{k,{\mathrm{ext}}})^{-1} + (\tilde{Q}_{k,{\mathrm{int}}})^{-1} } {2 \mathrm{i} (\omega - \tilde{\omega}_k)/ \tilde{\omega}_k + (\tilde{Q}_{k,{\mathrm{ext}}})^{-1} + (\tilde{Q}_{k,{\mathrm{int}}})^{-1}}
\label{ref_coe},
\end{equation} 
where $\tilde{\omega}_k$ are the corresponding excitation frequencies (with respect to the ground state) corresponding to the multi-particle eigenstates $\vert \tilde{\psi}_k \rangle$.
Note that the external and internal quality factors depend on the photonic fraction of the polariton component in such eigestates. This can be calculated by  weighing the eigenstate quality factors with their single-polariton fractions, namely:

\begin{equation}
    (\tilde{Q}_{k,\mathrm{int (ext)}})^{-1} = \sum_{k'> i_{0}} |\langle \tilde{\psi}_k \vert \hat{p}^{\dagger}_{k'} \vert 0 \rangle|^2 (\tilde{Q}^{pol}_{k',\mathrm{int (ext)}})^{-1}\,.
\end{equation}
The polariton quality factors depend on the photonic component of the polariton modes:

\begin{equation}
    (\tilde{Q}^{pol}_{k,\mathrm{int (ext)}})^{-1} = \sum_{i > i_{0}}   |W_{k ,i}|^2 (Q_{i,\mathrm{int (ext)}})^{-1} \, .
\end{equation}

Using Eq.~(\ref{ref_coe}), we can numerically generate the reflection coefficient of our multi-mode circuit QED system (see Fig.~\ref{S11}), which is shown in Fig.\ref{long_chain}. We used the values of $Q_{i=49,\textrm{int}} = 10000$ and $Q_{i=49,\textrm{ext}} = 2000$ to mimic a modest quality standing-wave resonance. The presence of a finite internal quality factor allows detecting the many-body resonances in the magnitude of the reflection amplitude $|S_{11}^{(s)}(\omega)|$. 

\section{Summary of device parameters}

The parameters used in our calculations are summarized in Table \ref{tableApp}.

\begin{center}
\begin{table}[h!]
\begin{tabular}{||l | r||}
 \hline
 Parameter &  Value  \\ [0ex] 
 \hline\hline
 Speed of light $v$ & $2.18 \times 10^{6} \: \mathrm{m/s}$ \\ 
 \hline
Wave impedance $Z_{\infty}$ & $9695 \: \mathrm{\Omega}$  \\
 \hline
 Length $\ell$ & 6 \: $\textrm{mm}$  \\
 \hline
 Charging energy $E_C/h$  & $5.69  \: \mathrm{GHz}$ \\ [1ex] 
 \hline
 Josephson energy $E_J/h$  & $8.12  \: \mathrm{GHz}$ \\ [1ex] 
 \hline
  Inductive energy $E_L/h$  & $1.42  \: \mathrm{GHz}$ \\ [1ex] 
 \hline
\end{tabular}
\caption{Device parameters used in our theoretical calculations. The inductive coupling ratio is $x = 1/2$.} 
\label{tableApp}
\end{table}
\end{center}

\bibliography{bib}

\begin{thebibliography}{57}%
\makeatletter
\providecommand \@ifxundefined [1]{%
 \@ifx{#1\undefined}
}%
\providecommand \@ifnum [1]{%
 \ifnum #1\expandafter \@firstoftwo
 \else \expandafter \@secondoftwo
 \fi
}%
\providecommand \@ifx [1]{%
 \ifx #1\expandafter \@firstoftwo
 \else \expandafter \@secondoftwo
 \fi
}%
\providecommand \natexlab [1]{#1}%
\providecommand \enquote  [1]{``#1''}%
\providecommand \bibnamefont  [1]{#1}%
\providecommand \bibfnamefont [1]{#1}%
\providecommand \citenamefont [1]{#1}%
\providecommand \href@noop [0]{\@secondoftwo}%
\providecommand \href [0]{\begingroup \@sanitize@url \@href}%
\providecommand \@href[1]{\@@startlink{#1}\@@href}%
\providecommand \@@href[1]{\endgroup#1\@@endlink}%
\providecommand \@sanitize@url [0]{\catcode `\\12\catcode `\$12\catcode
  `\&12\catcode `\#12\catcode `\^12\catcode `\_12\catcode `\%12\relax}%
\providecommand \@@startlink[1]{}%
\providecommand \@@endlink[0]{}%
\providecommand \url  [0]{\begingroup\@sanitize@url \@url }%
\providecommand \@url [1]{\endgroup\@href {#1}{\urlprefix }}%
\providecommand \urlprefix  [0]{URL }%
\providecommand \Eprint [0]{\href }%
\providecommand \doibase [0]{http://dx.doi.org/}%
\providecommand \selectlanguage [0]{\@gobble}%
\providecommand \bibinfo  [0]{\@secondoftwo}%
\providecommand \bibfield  [0]{\@secondoftwo}%
\providecommand \translation [1]{[#1]}%
\providecommand \BibitemOpen [0]{}%
\providecommand \bibitemStop [0]{}%
\providecommand \bibitemNoStop [0]{.\EOS\space}%
\providecommand \EOS [0]{\spacefactor3000\relax}%
\providecommand \BibitemShut  [1]{\csname bibitem#1\endcsname}%
\let\auto@bib@innerbib\@empty
\bibitem [{\citenamefont {Feynman}(1998)}]{Feynman_QED}%
  \BibitemOpen
  \bibfield  {author} {\bibinfo {author} {\bibfnamefont {R.~P.}\ \bibnamefont
  {Feynman}},\ }\href@noop {} {\emph {\bibinfo {title} {Quantum
  Electrodynamics}}}\ (\bibinfo  {publisher} {New Edition, Westview Press},\
  \bibinfo {year} {1998})\BibitemShut {NoStop}%
\bibitem [{\citenamefont {Milonni}(1994)}]{Milonni_1994}%
  \BibitemOpen
  \bibfield  {author} {\bibinfo {author} {\bibfnamefont {P.~W.}\ \bibnamefont
  {Milonni}},\ }\href@noop {} {\emph {\bibinfo {title} {The Quantum Vacuum}}}\
  (\bibinfo  {publisher} {Academic Press},\ \bibinfo {year} {1994})\BibitemShut
  {NoStop}%
\bibitem [{\citenamefont {Cohen-Tannoudji}\ \emph {et~al.}(1997)\citenamefont
  {Cohen-Tannoudji}, \citenamefont {Dupont-Roc},\ and\ \citenamefont
  {Grynberg}}]{Cohen_Photons}%
  \BibitemOpen
  \bibfield  {author} {\bibinfo {author} {\bibfnamefont {C.}~\bibnamefont
  {Cohen-Tannoudji}}, \bibinfo {author} {\bibfnamefont {J.}~\bibnamefont
  {Dupont-Roc}}, \ and\ \bibinfo {author} {\bibfnamefont {G.}~\bibnamefont
  {Grynberg}},\ }\href@noop {} {\emph {\bibinfo {title} {Photons and Atoms}}}\
  (\bibinfo  {publisher} {Wiley},\ \bibinfo {year} {1997})\BibitemShut
  {NoStop}%
\bibitem [{\citenamefont {Haroche}\ and\ \citenamefont
  {Raimond}(2006)}]{Haroche2006}%
  \BibitemOpen
  \bibfield  {author} {\bibinfo {author} {\bibfnamefont {S.}~\bibnamefont
  {Haroche}}\ and\ \bibinfo {author} {\bibfnamefont {J.-M.}\ \bibnamefont
  {Raimond}},\ }\href {\doibase 10.1093/acprof:oso/9780198509141.001.0001}
  {\emph {\bibinfo {title} {Exploring the Quantum}}}\ (\bibinfo  {publisher}
  {Oxford University Press},\ \bibinfo {year} {2006})\BibitemShut {NoStop}%
\bibitem [{ABC(2020)}]{ABC_CircuitQED}%
  \BibitemOpen
  \href {\doibase 10.1038/s41567-020-0847-3} {\bibfield  {journal} {\bibinfo
  {journal} {Nature Physics}\ }\textbf {\bibinfo {volume} {16}},\ \bibinfo
  {pages} {233} (\bibinfo {year} {2020})}\BibitemShut {NoStop}%
\bibitem [{\citenamefont {Blais}\ \emph {et~al.}(2021)\citenamefont {Blais},
  \citenamefont {Grimsmo}, \citenamefont {Girvin},\ and\ \citenamefont
  {Wallraff}}]{RMP_circuitQED}%
  \BibitemOpen
  \bibfield  {author} {\bibinfo {author} {\bibfnamefont {A.}~\bibnamefont
  {Blais}}, \bibinfo {author} {\bibfnamefont {A.~L.}\ \bibnamefont {Grimsmo}},
  \bibinfo {author} {\bibfnamefont {S.~M.}\ \bibnamefont {Girvin}}, \ and\
  \bibinfo {author} {\bibfnamefont {A.}~\bibnamefont {Wallraff}},\ }\href
  {\doibase 10.1103/RevModPhys.93.025005} {\bibfield  {journal} {\bibinfo
  {journal} {Rev. Mod. Phys.}\ }\textbf {\bibinfo {volume} {93}},\ \bibinfo
  {pages} {025005} (\bibinfo {year} {2021})}\BibitemShut {NoStop}%
\bibitem [{\citenamefont {Ciuti}\ \emph {et~al.}(2005)\citenamefont {Ciuti},
  \citenamefont {Bastard},\ and\ \citenamefont {Carusotto}}]{Ciuti_2005}%
  \BibitemOpen
  \bibfield  {author} {\bibinfo {author} {\bibfnamefont {C.}~\bibnamefont
  {Ciuti}}, \bibinfo {author} {\bibfnamefont {G.}~\bibnamefont {Bastard}}, \
  and\ \bibinfo {author} {\bibfnamefont {I.}~\bibnamefont {Carusotto}},\ }\href
  {\doibase 10.1103/PhysRevB.72.115303} {\bibfield  {journal} {\bibinfo
  {journal} {Phys. Rev. B}\ }\textbf {\bibinfo {volume} {72}},\ \bibinfo
  {pages} {115303} (\bibinfo {year} {2005})}\BibitemShut {NoStop}%
\bibitem [{\citenamefont {Forn-D\'{\i}az}\ \emph {et~al.}(2019)\citenamefont
  {Forn-D\'{\i}az}, \citenamefont {Lamata}, \citenamefont {Rico}, \citenamefont
  {Kono},\ and\ \citenamefont {Solano}}]{Forn_2019}%
  \BibitemOpen
  \bibfield  {author} {\bibinfo {author} {\bibfnamefont {P.}~\bibnamefont
  {Forn-D\'{\i}az}}, \bibinfo {author} {\bibfnamefont {L.}~\bibnamefont
  {Lamata}}, \bibinfo {author} {\bibfnamefont {E.}~\bibnamefont {Rico}},
  \bibinfo {author} {\bibfnamefont {J.}~\bibnamefont {Kono}}, \ and\ \bibinfo
  {author} {\bibfnamefont {E.}~\bibnamefont {Solano}},\ }\href {\doibase
  10.1103/RevModPhys.91.025005} {\bibfield  {journal} {\bibinfo  {journal}
  {Rev. Mod. Phys.}\ }\textbf {\bibinfo {volume} {91}},\ \bibinfo {pages}
  {025005} (\bibinfo {year} {2019})}\BibitemShut {NoStop}%
\bibitem [{\citenamefont {Kockum}\ \emph {et~al.}(2019)\citenamefont {Kockum},
  \citenamefont {Miranowicz}, \citenamefont {Liberato}, \citenamefont
  {Savasta},\ and\ \citenamefont {Nori}}]{Frisk_Kockum_2019}%
  \BibitemOpen
  \bibfield  {author} {\bibinfo {author} {\bibfnamefont {A.~F.}\ \bibnamefont
  {Kockum}}, \bibinfo {author} {\bibfnamefont {A.}~\bibnamefont {Miranowicz}},
  \bibinfo {author} {\bibfnamefont {S.~D.}\ \bibnamefont {Liberato}}, \bibinfo
  {author} {\bibfnamefont {S.}~\bibnamefont {Savasta}}, \ and\ \bibinfo
  {author} {\bibfnamefont {F.}~\bibnamefont {Nori}},\ }\href {\doibase
  10.1038/s42254-018-0006-2} {\bibfield  {journal} {\bibinfo  {journal} {Nature
  Reviews Physics}\ }\textbf {\bibinfo {volume} {1}},\ \bibinfo {pages} {19}
  (\bibinfo {year} {2019})}\BibitemShut {NoStop}%
\bibitem [{\citenamefont {Garcia-Vidal}\ \emph {et~al.}(2021)\citenamefont
  {Garcia-Vidal}, \citenamefont {Ciuti},\ and\ \citenamefont
  {Ebbesen}}]{GarciaVidal2021}%
  \BibitemOpen
  \bibfield  {author} {\bibinfo {author} {\bibfnamefont {F.~J.}\ \bibnamefont
  {Garcia-Vidal}}, \bibinfo {author} {\bibfnamefont {C.}~\bibnamefont {Ciuti}},
  \ and\ \bibinfo {author} {\bibfnamefont {T.~W.}\ \bibnamefont {Ebbesen}},\
  }\href {\doibase 10.1126/science.abd0336} {\bibfield  {journal} {\bibinfo
  {journal} {Science}\ }\textbf {\bibinfo {volume} {373}} (\bibinfo {year}
  {2021}),\ 10.1126/science.abd0336}\BibitemShut {NoStop}%
\bibitem [{\citenamefont {Todorov}\ \emph {et~al.}(2010)\citenamefont
  {Todorov}, \citenamefont {Andrews}, \citenamefont {Colombelli}, \citenamefont
  {De~Liberato}, \citenamefont {Ciuti}, \citenamefont {Klang}, \citenamefont
  {Strasser},\ and\ \citenamefont {Sirtori}}]{Todorov_2010}%
  \BibitemOpen
  \bibfield  {author} {\bibinfo {author} {\bibfnamefont {Y.}~\bibnamefont
  {Todorov}}, \bibinfo {author} {\bibfnamefont {A.~M.}\ \bibnamefont
  {Andrews}}, \bibinfo {author} {\bibfnamefont {R.}~\bibnamefont {Colombelli}},
  \bibinfo {author} {\bibfnamefont {S.}~\bibnamefont {De~Liberato}}, \bibinfo
  {author} {\bibfnamefont {C.}~\bibnamefont {Ciuti}}, \bibinfo {author}
  {\bibfnamefont {P.}~\bibnamefont {Klang}}, \bibinfo {author} {\bibfnamefont
  {G.}~\bibnamefont {Strasser}}, \ and\ \bibinfo {author} {\bibfnamefont
  {C.}~\bibnamefont {Sirtori}},\ }\href {\doibase
  10.1103/PhysRevLett.105.196402} {\bibfield  {journal} {\bibinfo  {journal}
  {Phys. Rev. Lett.}\ }\textbf {\bibinfo {volume} {105}},\ \bibinfo {pages}
  {196402} (\bibinfo {year} {2010})}\BibitemShut {NoStop}%
\bibitem [{\citenamefont {Scalari}\ \emph {et~al.}(2012)\citenamefont
  {Scalari}, \citenamefont {Maissen}, \citenamefont {Turcinkova}, \citenamefont
  {Hagenmuller}, \citenamefont {Liberato}, \citenamefont {Ciuti}, \citenamefont
  {Reichl}, \citenamefont {Schuh}, \citenamefont {Wegscheider}, \citenamefont
  {Beck},\ and\ \citenamefont {Faist}}]{Scalari_2012}%
  \BibitemOpen
  \bibfield  {author} {\bibinfo {author} {\bibfnamefont {G.}~\bibnamefont
  {Scalari}}, \bibinfo {author} {\bibfnamefont {C.}~\bibnamefont {Maissen}},
  \bibinfo {author} {\bibfnamefont {D.}~\bibnamefont {Turcinkova}}, \bibinfo
  {author} {\bibfnamefont {D.}~\bibnamefont {Hagenmuller}}, \bibinfo {author}
  {\bibfnamefont {S.~D.}\ \bibnamefont {Liberato}}, \bibinfo {author}
  {\bibfnamefont {C.}~\bibnamefont {Ciuti}}, \bibinfo {author} {\bibfnamefont
  {C.}~\bibnamefont {Reichl}}, \bibinfo {author} {\bibfnamefont
  {D.}~\bibnamefont {Schuh}}, \bibinfo {author} {\bibfnamefont
  {W.}~\bibnamefont {Wegscheider}}, \bibinfo {author} {\bibfnamefont
  {M.}~\bibnamefont {Beck}}, \ and\ \bibinfo {author} {\bibfnamefont
  {J.}~\bibnamefont {Faist}},\ }\href {\doibase 10.1126/science.1216022}
  {\bibfield  {journal} {\bibinfo  {journal} {Science}\ }\textbf {\bibinfo
  {volume} {335}},\ \bibinfo {pages} {1323} (\bibinfo {year}
  {2012})}\BibitemShut {NoStop}%
\bibitem [{\citenamefont {Niemczyk}\ \emph {et~al.}(2010)\citenamefont
  {Niemczyk}, \citenamefont {Deppe}, \citenamefont {Huebl}, \citenamefont
  {Menzel}, \citenamefont {Hocke}, \citenamefont {Schwarz}, \citenamefont
  {Garcia-Ripoll}, \citenamefont {Zueco}, \citenamefont {Hümmer},
  \citenamefont {Solano}, \citenamefont {Marx},\ and\ \citenamefont
  {Gross}}]{Niemczyk_2010}%
  \BibitemOpen
  \bibfield  {author} {\bibinfo {author} {\bibfnamefont {T.}~\bibnamefont
  {Niemczyk}}, \bibinfo {author} {\bibfnamefont {F.}~\bibnamefont {Deppe}},
  \bibinfo {author} {\bibfnamefont {H.}~\bibnamefont {Huebl}}, \bibinfo
  {author} {\bibfnamefont {E.~P.}\ \bibnamefont {Menzel}}, \bibinfo {author}
  {\bibfnamefont {F.}~\bibnamefont {Hocke}}, \bibinfo {author} {\bibfnamefont
  {M.~J.}\ \bibnamefont {Schwarz}}, \bibinfo {author} {\bibfnamefont {J.~J.}\
  \bibnamefont {Garcia-Ripoll}}, \bibinfo {author} {\bibfnamefont
  {D.}~\bibnamefont {Zueco}}, \bibinfo {author} {\bibfnamefont
  {T.}~\bibnamefont {Hümmer}}, \bibinfo {author} {\bibfnamefont
  {E.}~\bibnamefont {Solano}}, \bibinfo {author} {\bibfnamefont
  {A.}~\bibnamefont {Marx}}, \ and\ \bibinfo {author} {\bibfnamefont
  {R.}~\bibnamefont {Gross}},\ }\href {\doibase 10.1038/nphys1730} {\bibfield
  {journal} {\bibinfo  {journal} {Nature Physics}\ }\textbf {\bibinfo {volume}
  {6}},\ \bibinfo {pages} {772} (\bibinfo {year} {2010})}\BibitemShut {NoStop}%
\bibitem [{\citenamefont {Yoshihara}\ \emph {et~al.}(2016)\citenamefont
  {Yoshihara}, \citenamefont {Fuse}, \citenamefont {Ashhab}, \citenamefont
  {Kakuyanagi}, \citenamefont {Saito},\ and\ \citenamefont
  {Semba}}]{Yoshihara_2016}%
  \BibitemOpen
  \bibfield  {author} {\bibinfo {author} {\bibfnamefont {F.}~\bibnamefont
  {Yoshihara}}, \bibinfo {author} {\bibfnamefont {T.}~\bibnamefont {Fuse}},
  \bibinfo {author} {\bibfnamefont {S.}~\bibnamefont {Ashhab}}, \bibinfo
  {author} {\bibfnamefont {K.}~\bibnamefont {Kakuyanagi}}, \bibinfo {author}
  {\bibfnamefont {S.}~\bibnamefont {Saito}}, \ and\ \bibinfo {author}
  {\bibfnamefont {K.}~\bibnamefont {Semba}},\ }\href {\doibase
  10.1038/nphys3906} {\bibfield  {journal} {\bibinfo  {journal} {Nature
  Physics}\ }\textbf {\bibinfo {volume} {13}},\ \bibinfo {pages} {44} (\bibinfo
  {year} {2016})}\BibitemShut {NoStop}%
\bibitem [{\citenamefont {Meiser}\ and\ \citenamefont
  {Meystre}(2006)}]{Meiser_2006}%
  \BibitemOpen
  \bibfield  {author} {\bibinfo {author} {\bibfnamefont {D.}~\bibnamefont
  {Meiser}}\ and\ \bibinfo {author} {\bibfnamefont {P.}~\bibnamefont
  {Meystre}},\ }\href {\doibase 10.1103/PhysRevA.74.065801} {\bibfield
  {journal} {\bibinfo  {journal} {Phys. Rev. A}\ }\textbf {\bibinfo {volume}
  {74}},\ \bibinfo {pages} {065801} (\bibinfo {year} {2006})}\BibitemShut
  {NoStop}%
\bibitem [{\citenamefont {Sundaresan}\ \emph {et~al.}(2015)\citenamefont
  {Sundaresan}, \citenamefont {Liu}, \citenamefont {Sadri}, \citenamefont
  {Sz\ifmmode~\mbox{\H{o}}\else \H{o}\fi{}cs}, \citenamefont {Underwood},
  \citenamefont {Malekakhlagh}, \citenamefont {T\"ureci},\ and\ \citenamefont
  {Houck}}]{Houck_2015}%
  \BibitemOpen
  \bibfield  {author} {\bibinfo {author} {\bibfnamefont {N.~M.}\ \bibnamefont
  {Sundaresan}}, \bibinfo {author} {\bibfnamefont {Y.}~\bibnamefont {Liu}},
  \bibinfo {author} {\bibfnamefont {D.}~\bibnamefont {Sadri}}, \bibinfo
  {author} {\bibfnamefont {L.~J.}\ \bibnamefont {Sz\ifmmode~\mbox{\H{o}}\else
  \H{o}\fi{}cs}}, \bibinfo {author} {\bibfnamefont {D.~L.}\ \bibnamefont
  {Underwood}}, \bibinfo {author} {\bibfnamefont {M.}~\bibnamefont
  {Malekakhlagh}}, \bibinfo {author} {\bibfnamefont {H.~E.}\ \bibnamefont
  {T\"ureci}}, \ and\ \bibinfo {author} {\bibfnamefont {A.~A.}\ \bibnamefont
  {Houck}},\ }\href {\doibase 10.1103/PhysRevX.5.021035} {\bibfield  {journal}
  {\bibinfo  {journal} {Phys. Rev. X}\ }\textbf {\bibinfo {volume} {5}},\
  \bibinfo {pages} {021035} (\bibinfo {year} {2015})}\BibitemShut {NoStop}%
\bibitem [{\citenamefont {Johnson}\ \emph {et~al.}(2019)\citenamefont
  {Johnson}, \citenamefont {Blaha}, \citenamefont {Ulanov}, \citenamefont
  {Rauschenbeutel}, \citenamefont {Schneeweiss},\ and\ \citenamefont
  {Volz}}]{Volz2019}%
  \BibitemOpen
  \bibfield  {author} {\bibinfo {author} {\bibfnamefont {A.}~\bibnamefont
  {Johnson}}, \bibinfo {author} {\bibfnamefont {M.}~\bibnamefont {Blaha}},
  \bibinfo {author} {\bibfnamefont {A.~E.}\ \bibnamefont {Ulanov}}, \bibinfo
  {author} {\bibfnamefont {A.}~\bibnamefont {Rauschenbeutel}}, \bibinfo
  {author} {\bibfnamefont {P.}~\bibnamefont {Schneeweiss}}, \ and\ \bibinfo
  {author} {\bibfnamefont {J.}~\bibnamefont {Volz}},\ }\href {\doibase
  10.1103/PhysRevLett.123.243602} {\bibfield  {journal} {\bibinfo  {journal}
  {Phys. Rev. Lett.}\ }\textbf {\bibinfo {volume} {123}},\ \bibinfo {pages}
  {243602} (\bibinfo {year} {2019})}\BibitemShut {NoStop}%
\bibitem [{\citenamefont {Mart{\'{\i}}nez}\ \emph {et~al.}(2019)\citenamefont
  {Mart{\'{\i}}nez}, \citenamefont {L{\'{e}}ger}, \citenamefont {Gheeraert},
  \citenamefont {Dassonneville}, \citenamefont {Planat}, \citenamefont
  {Foroughi}, \citenamefont {Krupko}, \citenamefont {Buisson}, \citenamefont
  {Naud}, \citenamefont {Hasch-Guichard}, \citenamefont {Florens},
  \citenamefont {Snyman},\ and\ \citenamefont {Roch}}]{Puertas_Mart_nez_2019}%
  \BibitemOpen
  \bibfield  {author} {\bibinfo {author} {\bibfnamefont {J.~P.}\ \bibnamefont
  {Mart{\'{\i}}nez}}, \bibinfo {author} {\bibfnamefont {S.}~\bibnamefont
  {L{\'{e}}ger}}, \bibinfo {author} {\bibfnamefont {N.}~\bibnamefont
  {Gheeraert}}, \bibinfo {author} {\bibfnamefont {R.}~\bibnamefont
  {Dassonneville}}, \bibinfo {author} {\bibfnamefont {L.}~\bibnamefont
  {Planat}}, \bibinfo {author} {\bibfnamefont {F.}~\bibnamefont {Foroughi}},
  \bibinfo {author} {\bibfnamefont {Y.}~\bibnamefont {Krupko}}, \bibinfo
  {author} {\bibfnamefont {O.}~\bibnamefont {Buisson}}, \bibinfo {author}
  {\bibfnamefont {C.}~\bibnamefont {Naud}}, \bibinfo {author} {\bibfnamefont
  {W.}~\bibnamefont {Hasch-Guichard}}, \bibinfo {author} {\bibfnamefont
  {S.}~\bibnamefont {Florens}}, \bibinfo {author} {\bibfnamefont
  {I.}~\bibnamefont {Snyman}}, \ and\ \bibinfo {author} {\bibfnamefont
  {N.}~\bibnamefont {Roch}},\ }\href {\doibase 10.1038/s41534-018-0104-0}
  {\bibfield  {journal} {\bibinfo  {journal} {npj Quantum Information}\
  }\textbf {\bibinfo {volume} {5}} (\bibinfo {year} {2019}),\
  10.1038/s41534-018-0104-0}\BibitemShut {NoStop}%
\bibitem [{\citenamefont {Kuzmin}\ \emph
  {et~al.}(2019{\natexlab{a}})\citenamefont {Kuzmin}, \citenamefont {Mehta},
  \citenamefont {Grabon}, \citenamefont {Mencia},\ and\ \citenamefont
  {Manucharyan}}]{Kuzmin_2019}%
  \BibitemOpen
  \bibfield  {author} {\bibinfo {author} {\bibfnamefont {R.}~\bibnamefont
  {Kuzmin}}, \bibinfo {author} {\bibfnamefont {N.}~\bibnamefont {Mehta}},
  \bibinfo {author} {\bibfnamefont {N.}~\bibnamefont {Grabon}}, \bibinfo
  {author} {\bibfnamefont {R.}~\bibnamefont {Mencia}}, \ and\ \bibinfo {author}
  {\bibfnamefont {V.~E.}\ \bibnamefont {Manucharyan}},\ }\href {\doibase
  10.1038/s41534-019-0134-2} {\bibfield  {journal} {\bibinfo  {journal} {npj
  Quantum Information}\ }\textbf {\bibinfo {volume} {5}} (\bibinfo {year}
  {2019}{\natexlab{a}}),\ 10.1038/s41534-019-0134-2}\BibitemShut {NoStop}%
\bibitem [{\citenamefont {Bassani}\ \emph
  {et~al.}(1977{\natexlab{a}})\citenamefont {Bassani}, \citenamefont {Forney},\
  and\ \citenamefont {Quattropani}}]{Bassani_1977}%
  \BibitemOpen
  \bibfield  {author} {\bibinfo {author} {\bibfnamefont {F.}~\bibnamefont
  {Bassani}}, \bibinfo {author} {\bibfnamefont {J.~J.}\ \bibnamefont {Forney}},
  \ and\ \bibinfo {author} {\bibfnamefont {A.}~\bibnamefont {Quattropani}},\
  }\href {\doibase 10.1103/PhysRevLett.39.1070} {\bibfield  {journal} {\bibinfo
   {journal} {Phys. Rev. Lett.}\ }\textbf {\bibinfo {volume} {39}},\ \bibinfo
  {pages} {1070} (\bibinfo {year} {1977}{\natexlab{a}})}\BibitemShut {NoStop}%
\bibitem [{\citenamefont {Todorov}\ and\ \citenamefont
  {Sirtori}(2012)}]{Todorov_2012}%
  \BibitemOpen
  \bibfield  {author} {\bibinfo {author} {\bibfnamefont {Y.}~\bibnamefont
  {Todorov}}\ and\ \bibinfo {author} {\bibfnamefont {C.}~\bibnamefont
  {Sirtori}},\ }\href {\doibase 10.1103/PhysRevB.85.045304} {\bibfield
  {journal} {\bibinfo  {journal} {Phys. Rev. B}\ }\textbf {\bibinfo {volume}
  {85}},\ \bibinfo {pages} {045304} (\bibinfo {year} {2012})}\BibitemShut
  {NoStop}%
\bibitem [{\citenamefont {Manucharyan}\ \emph {et~al.}(2017)\citenamefont
  {Manucharyan}, \citenamefont {Baksic},\ and\ \citenamefont
  {Ciuti}}]{Manucharyan_2017}%
  \BibitemOpen
  \bibfield  {author} {\bibinfo {author} {\bibfnamefont {V.~E.}\ \bibnamefont
  {Manucharyan}}, \bibinfo {author} {\bibfnamefont {A.}~\bibnamefont {Baksic}},
  \ and\ \bibinfo {author} {\bibfnamefont {C.}~\bibnamefont {Ciuti}},\ }\href
  {\doibase 10.1088/1751-8121/aa6fbc} {\bibfield  {journal} {\bibinfo
  {journal} {Journal of Physics A: Mathematical and Theoretical}\ }\textbf
  {\bibinfo {volume} {50}},\ \bibinfo {pages} {294001} (\bibinfo {year}
  {2017})}\BibitemShut {NoStop}%
\bibitem [{\citenamefont {De~Bernardis}\ \emph {et~al.}(2018)\citenamefont
  {De~Bernardis}, \citenamefont {Pilar}, \citenamefont {Jaako}, \citenamefont
  {De~Liberato},\ and\ \citenamefont {Rabl}}]{De_Bernardis_gauge}%
  \BibitemOpen
  \bibfield  {author} {\bibinfo {author} {\bibfnamefont {D.}~\bibnamefont
  {De~Bernardis}}, \bibinfo {author} {\bibfnamefont {P.}~\bibnamefont {Pilar}},
  \bibinfo {author} {\bibfnamefont {T.}~\bibnamefont {Jaako}}, \bibinfo
  {author} {\bibfnamefont {S.}~\bibnamefont {De~Liberato}}, \ and\ \bibinfo
  {author} {\bibfnamefont {P.}~\bibnamefont {Rabl}},\ }\href {\doibase
  10.1103/PhysRevA.98.053819} {\bibfield  {journal} {\bibinfo  {journal} {Phys.
  Rev. A}\ }\textbf {\bibinfo {volume} {98}},\ \bibinfo {pages} {053819}
  (\bibinfo {year} {2018})}\BibitemShut {NoStop}%
\bibitem [{\citenamefont {Di~Stefano}\ \emph {et~al.}(2019)\citenamefont
  {Di~Stefano}, \citenamefont {Settineri}, \citenamefont {Macrì},
  \citenamefont {Garziano}, \citenamefont {Stassi}, \citenamefont {Savasta},\
  and\ \citenamefont {Nori}}]{Di_Stefano_2019}%
  \BibitemOpen
  \bibfield  {author} {\bibinfo {author} {\bibfnamefont {O.}~\bibnamefont
  {Di~Stefano}}, \bibinfo {author} {\bibfnamefont {A.}~\bibnamefont
  {Settineri}}, \bibinfo {author} {\bibfnamefont {V.}~\bibnamefont {Macrì}},
  \bibinfo {author} {\bibfnamefont {L.}~\bibnamefont {Garziano}}, \bibinfo
  {author} {\bibfnamefont {R.}~\bibnamefont {Stassi}}, \bibinfo {author}
  {\bibfnamefont {S.}~\bibnamefont {Savasta}}, \ and\ \bibinfo {author}
  {\bibfnamefont {F.}~\bibnamefont {Nori}},\ }\href {\doibase
  10.1038/s41567-019-0534-4} {\bibfield  {journal} {\bibinfo  {journal} {Nature
  Physics}\ }\textbf {\bibinfo {volume} {15}},\ \bibinfo {pages} {803}
  (\bibinfo {year} {2019})}\BibitemShut {NoStop}%
\bibitem [{\citenamefont {Stokes}\ and\ \citenamefont
  {Nazir}(2019)}]{stokes2019gauge}%
  \BibitemOpen
  \bibfield  {author} {\bibinfo {author} {\bibfnamefont {A.}~\bibnamefont
  {Stokes}}\ and\ \bibinfo {author} {\bibfnamefont {A.}~\bibnamefont {Nazir}},\
  }\href@noop {} {\bibfield  {journal} {\bibinfo  {journal} {Nature
  communications}\ }\textbf {\bibinfo {volume} {10}},\ \bibinfo {pages} {1}
  (\bibinfo {year} {2019})}\BibitemShut {NoStop}%
\bibitem [{\citenamefont {Roth}\ \emph {et~al.}(2019)\citenamefont {Roth},
  \citenamefont {Hassler},\ and\ \citenamefont
  {DiVincenzo}}]{Di_Vincenzo_2019}%
  \BibitemOpen
  \bibfield  {author} {\bibinfo {author} {\bibfnamefont {M.}~\bibnamefont
  {Roth}}, \bibinfo {author} {\bibfnamefont {F.}~\bibnamefont {Hassler}}, \
  and\ \bibinfo {author} {\bibfnamefont {D.~P.}\ \bibnamefont {DiVincenzo}},\
  }\href {\doibase 10.1103/PhysRevResearch.1.033128} {\bibfield  {journal}
  {\bibinfo  {journal} {Phys. Rev. Research}\ }\textbf {\bibinfo {volume}
  {1}},\ \bibinfo {pages} {033128} (\bibinfo {year} {2019})}\BibitemShut
  {NoStop}%
\bibitem [{\citenamefont {Stokes}\ and\ \citenamefont
  {Nazir}(2020{\natexlab{a}})}]{Nazir2020}%
  \BibitemOpen
  \bibfield  {author} {\bibinfo {author} {\bibfnamefont {A.}~\bibnamefont
  {Stokes}}\ and\ \bibinfo {author} {\bibfnamefont {A.}~\bibnamefont {Nazir}},\
  }\href@noop {} {\bibfield  {journal} {\bibinfo  {journal} {preprint
  arXiv:2005.06499}\ } (\bibinfo {year} {2020}{\natexlab{a}})}\BibitemShut
  {NoStop}%
\bibitem [{\citenamefont {Stokes}\ and\ \citenamefont
  {Nazir}(2020{\natexlab{b}})}]{Nazir2020b}%
  \BibitemOpen
  \bibfield  {author} {\bibinfo {author} {\bibfnamefont {A.}~\bibnamefont
  {Stokes}}\ and\ \bibinfo {author} {\bibfnamefont {A.}~\bibnamefont {Nazir}},\
  }\href@noop {} {\bibfield  {journal} {\bibinfo  {journal} {preprint
  arXiv:2009.10662}\ } (\bibinfo {year} {2020}{\natexlab{b}})}\BibitemShut
  {NoStop}%
\bibitem [{\citenamefont {Dmytruk}\ and\ \citenamefont
  {Schir\'o}(2021)}]{dmytruk2021gauge}%
  \BibitemOpen
  \bibfield  {author} {\bibinfo {author} {\bibfnamefont {O.}~\bibnamefont
  {Dmytruk}}\ and\ \bibinfo {author} {\bibfnamefont {M.}~\bibnamefont
  {Schir\'o}},\ }\href {\doibase 10.1103/PhysRevB.103.075131} {\bibfield
  {journal} {\bibinfo  {journal} {Phys. Rev. B}\ }\textbf {\bibinfo {volume}
  {103}},\ \bibinfo {pages} {075131} (\bibinfo {year} {2021})}\BibitemShut
  {NoStop}%
\bibitem [{\citenamefont {Savasta}\ \emph {et~al.}(2021)\citenamefont
  {Savasta}, \citenamefont {Di~Stefano}, \citenamefont {Settineri},
  \citenamefont {Zueco}, \citenamefont {Hughes},\ and\ \citenamefont
  {Nori}}]{Savasta2021}%
  \BibitemOpen
  \bibfield  {author} {\bibinfo {author} {\bibfnamefont {S.}~\bibnamefont
  {Savasta}}, \bibinfo {author} {\bibfnamefont {O.}~\bibnamefont {Di~Stefano}},
  \bibinfo {author} {\bibfnamefont {A.}~\bibnamefont {Settineri}}, \bibinfo
  {author} {\bibfnamefont {D.}~\bibnamefont {Zueco}}, \bibinfo {author}
  {\bibfnamefont {S.}~\bibnamefont {Hughes}}, \ and\ \bibinfo {author}
  {\bibfnamefont {F.}~\bibnamefont {Nori}},\ }\href {\doibase
  10.1103/PhysRevA.103.053703} {\bibfield  {journal} {\bibinfo  {journal}
  {Phys. Rev. A}\ }\textbf {\bibinfo {volume} {103}},\ \bibinfo {pages}
  {053703} (\bibinfo {year} {2021})}\BibitemShut {NoStop}%
\bibitem [{\citenamefont {Stokes}\ and\ \citenamefont
  {Nazir}(2021)}]{Nazir2021}%
  \BibitemOpen
  \bibfield  {author} {\bibinfo {author} {\bibfnamefont {A.}~\bibnamefont
  {Stokes}}\ and\ \bibinfo {author} {\bibfnamefont {A.}~\bibnamefont {Nazir}},\
  }\href {\doibase 10.1103/PhysRevResearch.3.013116} {\bibfield  {journal}
  {\bibinfo  {journal} {Phys. Rev. Research}\ }\textbf {\bibinfo {volume}
  {3}},\ \bibinfo {pages} {013116} (\bibinfo {year} {2021})}\BibitemShut
  {NoStop}%
\bibitem [{\citenamefont {Salmon}\ \emph {et~al.}(2022)\citenamefont {Salmon},
  \citenamefont {Gustin}, \citenamefont {Settineri}, \citenamefont {Stefano},
  \citenamefont {Zueco}, \citenamefont {Savasta}, \citenamefont {Nori},\ and\
  \citenamefont {Hughes}}]{Savasta2022}%
  \BibitemOpen
  \bibfield  {author} {\bibinfo {author} {\bibfnamefont {W.}~\bibnamefont
  {Salmon}}, \bibinfo {author} {\bibfnamefont {C.}~\bibnamefont {Gustin}},
  \bibinfo {author} {\bibfnamefont {A.}~\bibnamefont {Settineri}}, \bibinfo
  {author} {\bibfnamefont {O.~D.}\ \bibnamefont {Stefano}}, \bibinfo {author}
  {\bibfnamefont {D.}~\bibnamefont {Zueco}}, \bibinfo {author} {\bibfnamefont
  {S.}~\bibnamefont {Savasta}}, \bibinfo {author} {\bibfnamefont
  {F.}~\bibnamefont {Nori}}, \ and\ \bibinfo {author} {\bibfnamefont
  {S.}~\bibnamefont {Hughes}},\ }\href {\doibase doi:10.1515/nanoph-2021-0718}
  {\bibfield  {journal} {\bibinfo  {journal} {Nanophotonics}\ }\textbf
  {\bibinfo {volume} {11}},\ \bibinfo {pages} {1573} (\bibinfo {year}
  {2022})}\BibitemShut {NoStop}%
\bibitem [{\citenamefont {Gely}\ \emph {et~al.}(2017)\citenamefont {Gely},
  \citenamefont {Parra-Rodriguez}, \citenamefont {Bothner}, \citenamefont
  {Blanter}, \citenamefont {Bosman}, \citenamefont {Solano},\ and\
  \citenamefont {Steele}}]{Steele_2017}%
  \BibitemOpen
  \bibfield  {author} {\bibinfo {author} {\bibfnamefont {M.~F.}\ \bibnamefont
  {Gely}}, \bibinfo {author} {\bibfnamefont {A.}~\bibnamefont
  {Parra-Rodriguez}}, \bibinfo {author} {\bibfnamefont {D.}~\bibnamefont
  {Bothner}}, \bibinfo {author} {\bibfnamefont {Y.~M.}\ \bibnamefont
  {Blanter}}, \bibinfo {author} {\bibfnamefont {S.~J.}\ \bibnamefont {Bosman}},
  \bibinfo {author} {\bibfnamefont {E.}~\bibnamefont {Solano}}, \ and\ \bibinfo
  {author} {\bibfnamefont {G.~A.}\ \bibnamefont {Steele}},\ }\href {\doibase
  10.1103/PhysRevB.95.245115} {\bibfield  {journal} {\bibinfo  {journal} {Phys.
  Rev. B}\ }\textbf {\bibinfo {volume} {95}},\ \bibinfo {pages} {245115}
  (\bibinfo {year} {2017})}\BibitemShut {NoStop}%
\bibitem [{\citenamefont {Malekakhlagh}\ \emph {et~al.}(2017)\citenamefont
  {Malekakhlagh}, \citenamefont {Petrescu},\ and\ \citenamefont
  {T\"ureci}}]{Tureci_2017}%
  \BibitemOpen
  \bibfield  {author} {\bibinfo {author} {\bibfnamefont {M.}~\bibnamefont
  {Malekakhlagh}}, \bibinfo {author} {\bibfnamefont {A.}~\bibnamefont
  {Petrescu}}, \ and\ \bibinfo {author} {\bibfnamefont {H.~E.}\ \bibnamefont
  {T\"ureci}},\ }\href {\doibase 10.1103/PhysRevLett.119.073601} {\bibfield
  {journal} {\bibinfo  {journal} {Phys. Rev. Lett.}\ }\textbf {\bibinfo
  {volume} {119}},\ \bibinfo {pages} {073601} (\bibinfo {year}
  {2017})}\BibitemShut {NoStop}%
\bibitem [{\citenamefont {Sinha}\ \emph {et~al.}(2022)\citenamefont {Sinha},
  \citenamefont {Khan}, \citenamefont {C\"uce},\ and\ \citenamefont
  {T\"ureci}}]{Tureci2022}%
  \BibitemOpen
  \bibfield  {author} {\bibinfo {author} {\bibfnamefont {K.}~\bibnamefont
  {Sinha}}, \bibinfo {author} {\bibfnamefont {S.~A.}\ \bibnamefont {Khan}},
  \bibinfo {author} {\bibfnamefont {E.}~\bibnamefont {C\"uce}}, \ and\ \bibinfo
  {author} {\bibfnamefont {H.~E.}\ \bibnamefont {T\"ureci}},\ }\href {\doibase
  10.1103/PhysRevA.106.033714} {\bibfield  {journal} {\bibinfo  {journal}
  {Phys. Rev. A}\ }\textbf {\bibinfo {volume} {106}},\ \bibinfo {pages}
  {033714} (\bibinfo {year} {2022})}\BibitemShut {NoStop}%
\bibitem [{\citenamefont {Koch}\ \emph {et~al.}(2007)\citenamefont {Koch},
  \citenamefont {Terri}, \citenamefont {Gambetta}, \citenamefont {Houck},
  \citenamefont {Schuster}, \citenamefont {Majer}, \citenamefont {Blais},
  \citenamefont {Devoret}, \citenamefont {Girvin},\ and\ \citenamefont
  {Schoelkopf}}]{koch2007charge}%
  \BibitemOpen
  \bibfield  {author} {\bibinfo {author} {\bibfnamefont {J.}~\bibnamefont
  {Koch}}, \bibinfo {author} {\bibfnamefont {M.~Y.}\ \bibnamefont {Terri}},
  \bibinfo {author} {\bibfnamefont {J.}~\bibnamefont {Gambetta}}, \bibinfo
  {author} {\bibfnamefont {A.~A.}\ \bibnamefont {Houck}}, \bibinfo {author}
  {\bibfnamefont {D.~I.}\ \bibnamefont {Schuster}}, \bibinfo {author}
  {\bibfnamefont {J.}~\bibnamefont {Majer}}, \bibinfo {author} {\bibfnamefont
  {A.}~\bibnamefont {Blais}}, \bibinfo {author} {\bibfnamefont {M.~H.}\
  \bibnamefont {Devoret}}, \bibinfo {author} {\bibfnamefont {S.~M.}\
  \bibnamefont {Girvin}}, \ and\ \bibinfo {author} {\bibfnamefont {R.~J.}\
  \bibnamefont {Schoelkopf}},\ }\href@noop {} {\bibfield  {journal} {\bibinfo
  {journal} {Physical Review A}\ }\textbf {\bibinfo {volume} {76}},\ \bibinfo
  {pages} {042319} (\bibinfo {year} {2007})}\BibitemShut {NoStop}%
\bibitem [{\citenamefont {Manucharyan}\ \emph {et~al.}(2007)\citenamefont
  {Manucharyan}, \citenamefont {Boaknin}, \citenamefont {Metcalfe},
  \citenamefont {Vijay}, \citenamefont {Siddiqi},\ and\ \citenamefont
  {Devoret}}]{manucharyan2007microwave}%
  \BibitemOpen
  \bibfield  {author} {\bibinfo {author} {\bibfnamefont {V.}~\bibnamefont
  {Manucharyan}}, \bibinfo {author} {\bibfnamefont {E.}~\bibnamefont
  {Boaknin}}, \bibinfo {author} {\bibfnamefont {M.}~\bibnamefont {Metcalfe}},
  \bibinfo {author} {\bibfnamefont {R.}~\bibnamefont {Vijay}}, \bibinfo
  {author} {\bibfnamefont {I.}~\bibnamefont {Siddiqi}}, \ and\ \bibinfo
  {author} {\bibfnamefont {M.}~\bibnamefont {Devoret}},\ }\href@noop {}
  {\bibfield  {journal} {\bibinfo  {journal} {Physical Review B}\ }\textbf
  {\bibinfo {volume} {76}},\ \bibinfo {pages} {014524} (\bibinfo {year}
  {2007})}\BibitemShut {NoStop}%
\bibitem [{\citenamefont {Nigg}\ \emph {et~al.}(2012)\citenamefont {Nigg},
  \citenamefont {Paik}, \citenamefont {Vlastakis}, \citenamefont {Kirchmair},
  \citenamefont {Shankar}, \citenamefont {Frunzio}, \citenamefont {Devoret},
  \citenamefont {Schoelkopf},\ and\ \citenamefont {Girvin}}]{Nigg_2012}%
  \BibitemOpen
  \bibfield  {author} {\bibinfo {author} {\bibfnamefont {S.~E.}\ \bibnamefont
  {Nigg}}, \bibinfo {author} {\bibfnamefont {H.}~\bibnamefont {Paik}}, \bibinfo
  {author} {\bibfnamefont {B.}~\bibnamefont {Vlastakis}}, \bibinfo {author}
  {\bibfnamefont {G.}~\bibnamefont {Kirchmair}}, \bibinfo {author}
  {\bibfnamefont {S.}~\bibnamefont {Shankar}}, \bibinfo {author} {\bibfnamefont
  {L.}~\bibnamefont {Frunzio}}, \bibinfo {author} {\bibfnamefont {M.~H.}\
  \bibnamefont {Devoret}}, \bibinfo {author} {\bibfnamefont {R.~J.}\
  \bibnamefont {Schoelkopf}}, \ and\ \bibinfo {author} {\bibfnamefont {S.~M.}\
  \bibnamefont {Girvin}},\ }\href {\doibase 10.1103/PhysRevLett.108.240502}
  {\bibfield  {journal} {\bibinfo  {journal} {Phys. Rev. Lett.}\ }\textbf
  {\bibinfo {volume} {108}},\ \bibinfo {pages} {240502} (\bibinfo {year}
  {2012})}\BibitemShut {NoStop}%
\bibitem [{\citenamefont {Manucharyan}\ \emph {et~al.}(2009)\citenamefont
  {Manucharyan}, \citenamefont {Koch}, \citenamefont {Glazman},\ and\
  \citenamefont {Devoret}}]{manucharyan2009fluxonium}%
  \BibitemOpen
  \bibfield  {author} {\bibinfo {author} {\bibfnamefont {V.~E.}\ \bibnamefont
  {Manucharyan}}, \bibinfo {author} {\bibfnamefont {J.}~\bibnamefont {Koch}},
  \bibinfo {author} {\bibfnamefont {L.~I.}\ \bibnamefont {Glazman}}, \ and\
  \bibinfo {author} {\bibfnamefont {M.~H.}\ \bibnamefont {Devoret}},\
  }\href@noop {} {\bibfield  {journal} {\bibinfo  {journal} {Science}\ }\textbf
  {\bibinfo {volume} {326}},\ \bibinfo {pages} {113} (\bibinfo {year}
  {2009})}\BibitemShut {NoStop}%
\bibitem [{\citenamefont {Manucharyan}\ \emph {et~al.}(2012)\citenamefont
  {Manucharyan}, \citenamefont {Masluk}, \citenamefont {Kamal}, \citenamefont
  {Koch}, \citenamefont {Glazman},\ and\ \citenamefont
  {Devoret}}]{manucharyan2012evidence}%
  \BibitemOpen
  \bibfield  {author} {\bibinfo {author} {\bibfnamefont {V.~E.}\ \bibnamefont
  {Manucharyan}}, \bibinfo {author} {\bibfnamefont {N.~A.}\ \bibnamefont
  {Masluk}}, \bibinfo {author} {\bibfnamefont {A.}~\bibnamefont {Kamal}},
  \bibinfo {author} {\bibfnamefont {J.}~\bibnamefont {Koch}}, \bibinfo {author}
  {\bibfnamefont {L.~I.}\ \bibnamefont {Glazman}}, \ and\ \bibinfo {author}
  {\bibfnamefont {M.~H.}\ \bibnamefont {Devoret}},\ }\href@noop {} {\bibfield
  {journal} {\bibinfo  {journal} {Physical Review B}\ }\textbf {\bibinfo
  {volume} {85}},\ \bibinfo {pages} {024521} (\bibinfo {year}
  {2012})}\BibitemShut {NoStop}%
\bibitem [{\citenamefont {Girvin}\ and\ \citenamefont
  {Yang}(2019)}]{Girvin_2019}%
  \BibitemOpen
  \bibfield  {author} {\bibinfo {author} {\bibfnamefont {S.~M.}\ \bibnamefont
  {Girvin}}\ and\ \bibinfo {author} {\bibfnamefont {K.}~\bibnamefont {Yang}},\
  }\href@noop {} {\emph {\bibinfo {title} {Modern Condensed Matter Physics}}}\
  (\bibinfo  {publisher} {Cambridge University Press},\ \bibinfo {year}
  {2019})\BibitemShut {NoStop}%
\bibitem [{\citenamefont {Forn-D{\'\i}az}\ \emph {et~al.}(2017)\citenamefont
  {Forn-D{\'\i}az}, \citenamefont {Garc{\'\i}a-Ripoll}, \citenamefont
  {Peropadre}, \citenamefont {Orgiazzi}, \citenamefont {Yurtalan},
  \citenamefont {Belyansky}, \citenamefont {Wilson},\ and\ \citenamefont
  {Lupascu}}]{forn2017ultrastrong}%
  \BibitemOpen
  \bibfield  {author} {\bibinfo {author} {\bibfnamefont {P.}~\bibnamefont
  {Forn-D{\'\i}az}}, \bibinfo {author} {\bibfnamefont {J.~J.}\ \bibnamefont
  {Garc{\'\i}a-Ripoll}}, \bibinfo {author} {\bibfnamefont {B.}~\bibnamefont
  {Peropadre}}, \bibinfo {author} {\bibfnamefont {J.-L.}\ \bibnamefont
  {Orgiazzi}}, \bibinfo {author} {\bibfnamefont {M.}~\bibnamefont {Yurtalan}},
  \bibinfo {author} {\bibfnamefont {R.}~\bibnamefont {Belyansky}}, \bibinfo
  {author} {\bibfnamefont {C.~M.}\ \bibnamefont {Wilson}}, \ and\ \bibinfo
  {author} {\bibfnamefont {A.}~\bibnamefont {Lupascu}},\ }\href@noop {}
  {\bibfield  {journal} {\bibinfo  {journal} {Nature Physics}\ }\textbf
  {\bibinfo {volume} {13}},\ \bibinfo {pages} {39} (\bibinfo {year}
  {2017})}\BibitemShut {NoStop}%
\bibitem [{\citenamefont {Goldstein}\ \emph {et~al.}(2013)\citenamefont
  {Goldstein}, \citenamefont {Devoret}, \citenamefont {Houzet},\ and\
  \citenamefont {Glazman}}]{Goldstein_2013}%
  \BibitemOpen
  \bibfield  {author} {\bibinfo {author} {\bibfnamefont {M.}~\bibnamefont
  {Goldstein}}, \bibinfo {author} {\bibfnamefont {M.~H.}\ \bibnamefont
  {Devoret}}, \bibinfo {author} {\bibfnamefont {M.}~\bibnamefont {Houzet}}, \
  and\ \bibinfo {author} {\bibfnamefont {L.~I.}\ \bibnamefont {Glazman}},\
  }\href {\doibase 10.1103/PhysRevLett.110.017002} {\bibfield  {journal}
  {\bibinfo  {journal} {Phys. Rev. Lett.}\ }\textbf {\bibinfo {volume} {110}},\
  \bibinfo {pages} {017002} (\bibinfo {year} {2013})}\BibitemShut {NoStop}%
\bibitem [{\citenamefont {Gheeraert}\ \emph {et~al.}(2018)\citenamefont
  {Gheeraert}, \citenamefont {Zhang}, \citenamefont {S\'epulcre}, \citenamefont
  {Bera}, \citenamefont {Roch}, \citenamefont {Baranger},\ and\ \citenamefont
  {Florens}}]{Florens_2018}%
  \BibitemOpen
  \bibfield  {author} {\bibinfo {author} {\bibfnamefont {N.}~\bibnamefont
  {Gheeraert}}, \bibinfo {author} {\bibfnamefont {X.~H.~H.}\ \bibnamefont
  {Zhang}}, \bibinfo {author} {\bibfnamefont {T.}~\bibnamefont {S\'epulcre}},
  \bibinfo {author} {\bibfnamefont {S.}~\bibnamefont {Bera}}, \bibinfo {author}
  {\bibfnamefont {N.}~\bibnamefont {Roch}}, \bibinfo {author} {\bibfnamefont
  {H.~U.}\ \bibnamefont {Baranger}}, \ and\ \bibinfo {author} {\bibfnamefont
  {S.}~\bibnamefont {Florens}},\ }\href {\doibase 10.1103/PhysRevA.98.043816}
  {\bibfield  {journal} {\bibinfo  {journal} {Phys. Rev. A}\ }\textbf {\bibinfo
  {volume} {98}},\ \bibinfo {pages} {043816} (\bibinfo {year}
  {2018})}\BibitemShut {NoStop}%
\bibitem [{\citenamefont {Houzet}\ and\ \citenamefont
  {Glazman}(2020)}]{houzet2020critical}%
  \BibitemOpen
  \bibfield  {author} {\bibinfo {author} {\bibfnamefont {M.}~\bibnamefont
  {Houzet}}\ and\ \bibinfo {author} {\bibfnamefont {L.}~\bibnamefont
  {Glazman}},\ }\href@noop {} {\bibfield  {journal} {\bibinfo  {journal}
  {Physical Review Letters}\ }\textbf {\bibinfo {volume} {125}},\ \bibinfo
  {pages} {267701} (\bibinfo {year} {2020})}\BibitemShut {NoStop}%
\bibitem [{\citenamefont {Burshtein}\ \emph {et~al.}(2021)\citenamefont
  {Burshtein}, \citenamefont {Kuzmin}, \citenamefont {Manucharyan},\ and\
  \citenamefont {Goldstein}}]{burshtein2021photon}%
  \BibitemOpen
  \bibfield  {author} {\bibinfo {author} {\bibfnamefont {A.}~\bibnamefont
  {Burshtein}}, \bibinfo {author} {\bibfnamefont {R.}~\bibnamefont {Kuzmin}},
  \bibinfo {author} {\bibfnamefont {V.~E.}\ \bibnamefont {Manucharyan}}, \ and\
  \bibinfo {author} {\bibfnamefont {M.}~\bibnamefont {Goldstein}},\ }\href@noop
  {} {\bibfield  {journal} {\bibinfo  {journal} {Physical Review Letters}\
  }\textbf {\bibinfo {volume} {126}},\ \bibinfo {pages} {137701} (\bibinfo
  {year} {2021})}\BibitemShut {NoStop}%
\bibitem [{\citenamefont {Klyshko}(1969)}]{klyshko1969scattering}%
  \BibitemOpen
  \bibfield  {author} {\bibinfo {author} {\bibfnamefont {D.}~\bibnamefont
  {Klyshko}},\ }\href {http://www.jetp.ras.ru/cgi-bin/dn/e_028_03_0522.pdf}
  {\bibfield  {journal} {\bibinfo  {journal} {Sov. Phys. JETP}\ }\textbf
  {\bibinfo {volume} {28}},\ \bibinfo {pages} {522} (\bibinfo {year}
  {1969})}\BibitemShut {NoStop}%
\bibitem [{\citenamefont {Harris}\ \emph {et~al.}(1967)\citenamefont {Harris},
  \citenamefont {Oshman},\ and\ \citenamefont {Byer}}]{harris1967observation}%
  \BibitemOpen
  \bibfield  {author} {\bibinfo {author} {\bibfnamefont {S.}~\bibnamefont
  {Harris}}, \bibinfo {author} {\bibfnamefont {M.}~\bibnamefont {Oshman}}, \
  and\ \bibinfo {author} {\bibfnamefont {R.}~\bibnamefont {Byer}},\ }\href@noop
  {} {\bibfield  {journal} {\bibinfo  {journal} {Physical Review Letters}\
  }\textbf {\bibinfo {volume} {18}},\ \bibinfo {pages} {732} (\bibinfo {year}
  {1967})}\BibitemShut {NoStop}%
\bibitem [{\citenamefont {Yanagimoto}\ \emph {et~al.}(2022)\citenamefont
  {Yanagimoto}, \citenamefont {Ng}, \citenamefont {Jankowski}, \citenamefont
  {Mabuchi},\ and\ \citenamefont {Hamerly}}]{yanagimoto2022temporal}%
  \BibitemOpen
  \bibfield  {author} {\bibinfo {author} {\bibfnamefont {R.}~\bibnamefont
  {Yanagimoto}}, \bibinfo {author} {\bibfnamefont {E.}~\bibnamefont {Ng}},
  \bibinfo {author} {\bibfnamefont {M.}~\bibnamefont {Jankowski}}, \bibinfo
  {author} {\bibfnamefont {H.}~\bibnamefont {Mabuchi}}, \ and\ \bibinfo
  {author} {\bibfnamefont {R.}~\bibnamefont {Hamerly}},\ }\href
  {https://arxiv.org/abs/2203.11909} {\bibfield  {journal} {\bibinfo  {journal}
  {arXiv preprint arXiv:2203.11909}\ } (\bibinfo {year} {2022})}\BibitemShut
  {NoStop}%
\bibitem [{\citenamefont {Bassani}\ \emph
  {et~al.}(1977{\natexlab{b}})\citenamefont {Bassani}, \citenamefont {Forney},\
  and\ \citenamefont {Quattropani}}]{Bassani1977}%
  \BibitemOpen
  \bibfield  {author} {\bibinfo {author} {\bibfnamefont {F.}~\bibnamefont
  {Bassani}}, \bibinfo {author} {\bibfnamefont {J.~J.}\ \bibnamefont {Forney}},
  \ and\ \bibinfo {author} {\bibfnamefont {A.}~\bibnamefont {Quattropani}},\
  }\href {\doibase 10.1103/PhysRevLett.39.1070} {\bibfield  {journal} {\bibinfo
   {journal} {Phys. Rev. Lett.}\ }\textbf {\bibinfo {volume} {39}},\ \bibinfo
  {pages} {1070} (\bibinfo {year} {1977}{\natexlab{b}})}\BibitemShut {NoStop}%
\bibitem [{\citenamefont {Koshino}\ \emph {et~al.}(2022)\citenamefont
  {Koshino}, \citenamefont {Shitara}, \citenamefont {Ao},\ and\ \citenamefont
  {Semba}}]{Koshino2022}%
  \BibitemOpen
  \bibfield  {author} {\bibinfo {author} {\bibfnamefont {K.}~\bibnamefont
  {Koshino}}, \bibinfo {author} {\bibfnamefont {T.}~\bibnamefont {Shitara}},
  \bibinfo {author} {\bibfnamefont {Z.}~\bibnamefont {Ao}}, \ and\ \bibinfo
  {author} {\bibfnamefont {K.}~\bibnamefont {Semba}},\ }\href {\doibase
  10.1103/PhysRevResearch.4.013013} {\bibfield  {journal} {\bibinfo  {journal}
  {Phys. Rev. Research}\ }\textbf {\bibinfo {volume} {4}},\ \bibinfo {pages}
  {013013} (\bibinfo {year} {2022})}\BibitemShut {NoStop}%
\bibitem [{\citenamefont {Altshuler}\ \emph {et~al.}(1997)\citenamefont
  {Altshuler}, \citenamefont {Gefen}, \citenamefont {Kamenev},\ and\
  \citenamefont {Levitov}}]{Altschuler1997}%
  \BibitemOpen
  \bibfield  {author} {\bibinfo {author} {\bibfnamefont {B.~L.}\ \bibnamefont
  {Altshuler}}, \bibinfo {author} {\bibfnamefont {Y.}~\bibnamefont {Gefen}},
  \bibinfo {author} {\bibfnamefont {A.}~\bibnamefont {Kamenev}}, \ and\
  \bibinfo {author} {\bibfnamefont {L.~S.}\ \bibnamefont {Levitov}},\ }\href
  {\doibase 10.1103/PhysRevLett.78.2803} {\bibfield  {journal} {\bibinfo
  {journal} {Phys. Rev. Lett.}\ }\textbf {\bibinfo {volume} {78}},\ \bibinfo
  {pages} {2803} (\bibinfo {year} {1997})}\BibitemShut {NoStop}%
\bibitem [{\citenamefont {Kuzmin}\ \emph {et~al.}(2021)\citenamefont {Kuzmin},
  \citenamefont {Grabon}, \citenamefont {Mehta}, \citenamefont {Burshtein},
  \citenamefont {Goldstein}, \citenamefont {Houzet}, \citenamefont {Glazman},\
  and\ \citenamefont {Manucharyan}}]{Kuzmin_2021}%
  \BibitemOpen
  \bibfield  {author} {\bibinfo {author} {\bibfnamefont {R.}~\bibnamefont
  {Kuzmin}}, \bibinfo {author} {\bibfnamefont {N.}~\bibnamefont {Grabon}},
  \bibinfo {author} {\bibfnamefont {N.}~\bibnamefont {Mehta}}, \bibinfo
  {author} {\bibfnamefont {A.}~\bibnamefont {Burshtein}}, \bibinfo {author}
  {\bibfnamefont {M.}~\bibnamefont {Goldstein}}, \bibinfo {author}
  {\bibfnamefont {M.}~\bibnamefont {Houzet}}, \bibinfo {author} {\bibfnamefont
  {L.~I.}\ \bibnamefont {Glazman}}, \ and\ \bibinfo {author} {\bibfnamefont
  {V.~E.}\ \bibnamefont {Manucharyan}},\ }\href {\doibase
  10.1103/PhysRevLett.126.197701} {\bibfield  {journal} {\bibinfo  {journal}
  {Phys. Rev. Lett.}\ }\textbf {\bibinfo {volume} {126}},\ \bibinfo {pages}
  {197701} (\bibinfo {year} {2021})}\BibitemShut {NoStop}%
\bibitem [{\citenamefont {N.}\ \emph {et~al.}(2022)\citenamefont {N.},
  \citenamefont {Mehta}, \citenamefont {Kuzmin}, \citenamefont {Ciuti},\ and\
  \citenamefont {Manucharyan}}]{Nitish_Nature}%
  \BibitemOpen
  \bibfield  {author} {\bibinfo {author} {\bibnamefont {N.}}, \bibinfo {author}
  {\bibnamefont {Mehta}}, \bibinfo {author} {\bibfnamefont {R.}~\bibnamefont
  {Kuzmin}}, \bibinfo {author} {\bibfnamefont {C.}~\bibnamefont {Ciuti}}, \
  and\ \bibinfo {author} {\bibfnamefont {V.~E.}\ \bibnamefont {Manucharyan}},\
  }\href {https://doi.org/10.48550/arXiv.2203.17186} {\  (\bibinfo {year}
  {2022})},\ \Eprint {http://arxiv.org/abs/2203.17186} {arXiv:2203.17186}
  \BibitemShut {NoStop}%
\bibitem [{\citenamefont {{Foster}}(1924)}]{foster_1924}%
  \BibitemOpen
  \bibfield  {author} {\bibinfo {author} {\bibfnamefont {R.~M.}\ \bibnamefont
  {{Foster}}},\ }\href {\doibase 10.1002/j.1538-7305.1924.tb01358.x} {\bibfield
   {journal} {\bibinfo  {journal} {The Bell System Technical Journal}\ }\textbf
  {\bibinfo {volume} {3}},\ \bibinfo {pages} {259} (\bibinfo {year}
  {1924})}\BibitemShut {NoStop}%
\bibitem [{\citenamefont {Pozar}(2005)}]{Pozar}%
  \BibitemOpen
  \bibfield  {author} {\bibinfo {author} {\bibfnamefont {D.~M.}\ \bibnamefont
  {Pozar}},\ }\href {https://cds.cern.ch/record/882338} {\emph {\bibinfo
  {title} {{Microwave engineering; 3rd ed.}}}}\ (\bibinfo  {publisher}
  {Wiley},\ \bibinfo {address} {Hoboken, NJ},\ \bibinfo {year}
  {2005})\BibitemShut {NoStop}%
\bibitem [{\citenamefont {Kuzmin}\ \emph
  {et~al.}(2019{\natexlab{b}})\citenamefont {Kuzmin}, \citenamefont {Mencia},
  \citenamefont {Grabon}, \citenamefont {Mehta}, \citenamefont {Lin},\ and\
  \citenamefont {Manucharyan}}]{Kuzmin2019_1}%
  \BibitemOpen
  \bibfield  {author} {\bibinfo {author} {\bibfnamefont {R.}~\bibnamefont
  {Kuzmin}}, \bibinfo {author} {\bibfnamefont {R.}~\bibnamefont {Mencia}},
  \bibinfo {author} {\bibfnamefont {N.}~\bibnamefont {Grabon}}, \bibinfo
  {author} {\bibfnamefont {N.}~\bibnamefont {Mehta}}, \bibinfo {author}
  {\bibfnamefont {Y.~H.}\ \bibnamefont {Lin}}, \ and\ \bibinfo {author}
  {\bibfnamefont {V.~E.}\ \bibnamefont {Manucharyan}},\ }\bibfield  {booktitle}
  {\emph {\bibinfo {booktitle} {Nature Physics}},\ }\href {\doibase
  10.1038/s41567-019-0553-1} {\ \textbf {\bibinfo {volume} {15}},\ \bibinfo
  {pages} {930} (\bibinfo {year} {2019}{\natexlab{b}})},\ \Eprint
  {http://arxiv.org/abs/1805.07379} {arXiv:1805.07379} \BibitemShut {NoStop}%
\end{thebibliography}%
\bibliographystyle{apsrev4-1}
\clearpage

\end{document}